\documentclass[%
 reprint,
 superscriptaddress,
 amsmath,amssymb,
]{revtex4-2}

\usepackage{graphicx}
\usepackage{dcolumn}
\usepackage{bm}
\usepackage{hyperref}
\hypersetup{
    colorlinks=true,
    linkcolor=blue,
    filecolor=blue,
    urlcolor=blue,
    citecolor=blue,
}
\usepackage{siunitx}

\usepackage{eso-pic}
\begin{document} 
\title{Experimental demonstration of a tomographic 5D phase-space reconstruction}
 
\author{S. Jaster-Merz}
\email[]{sonja.jaster-merz@desy.de} 

\author{R. W. Assmann}
\thanks{Present address: GSI Helmholtzzentrum für Schwerionenforschung GmbH, Darmstadt, Germany} 
\author{J. Beinortait\.e}
\author{J. Bj\"{o}rklund Svensson}
\thanks{Present address: Lund University, Lund, Sweden}
\author{R. Brinkmann}
\author{F. Burkart}
\affiliation{Deutsches Elektronen-Synchrotron DESY, Germany}
\author{P. Craievich}
\affiliation{Paul Scherrer Institut, Switzerland}
\author{H. Dinter}
\author{P. Gonz\'alez Caminal}
\thanks{Present address: ATG Science \& Technology S.L., Barcelona, Spain}
\affiliation{Deutsches Elektronen-Synchrotron DESY, Germany}
\author{W. Hillert}
\affiliation{Department of Physics Universität Hamburg, Germany}
\author{A. L. Kanekar}
\author{M. Kellermeier}
\author{W. Kuropka}
\author{F. Mayet}
\author{J. Osterhoff}
\thanks{Present address: Lawrence Berkeley National Laboratory, Berkeley, California 94720, USA}
\author{B. Stacey}
\author{M. Stanitzki}
\author{T. Vinatier}
\author{S. Wesch}
\author{R. D'Arcy}
\thanks{Present address: Oxford University, Oxford, United Kingdom}
\affiliation{Deutsches Elektronen-Synchrotron DESY, Germany}

\date{\today}

\begin{abstract}
    Detailed knowledge of particle-beam properties is of great importance to understand and push the performance of existing and next-generation particle accelerators. We recently proposed a new phase-space tomography method to reconstruct the five-dimensional (5D) phase space, i.e., the charge density distribution in all three spatial directions and the two transverse momenta. 
    Here, we present the first experimental demonstration of the method at the FLASHForward facility at DESY. This includes the reconstruction of the 5D phase-space distribution of a GeV-class electron bunch, the use of this measured phase space to create a particle distribution for simulations, and the extraction of the transverse 4D slice emittance. 
\end{abstract}

\maketitle

Particle accelerators are highly complex and multivariate machines that are required to produce beams with percent- to sub-permille-level stability at highly specialized working points with limited diagnostics. 
The generation of these beams is especially difficult for applications that require excellent phase-space quality such as advanced ultra-high-gradient accelerators \cite{PhysRevLett.54.693, PhysRevLett.67.991, PhysRevLett.129.234801, Joshi_2018, Adli2018, Blumenfeld2007, Litos2014, FERRARIO2013183, Pompili2022, photonics10020099} or free-electron lasers \cite{ROSSBACH20191, Abela:77248, Emma2010, Ishikawa2012, giannessi:fel2019-thp079, Kang2017, Prat2020, liu2021sxfel}, and their complex operation modes (e.g. two-color lasing \cite{Ferrari2016, PhysRevResearch.4.L022025}).

To characterize these electron beams, methods focussing on the reconstruction of statistical beam parameters or 2-dimensional (2D) projections of the full phase space \cite{MCKEE1995264, PhysRevSTAB.9.112801, PhysRevSTAB.6.122801, PhysRevAccelBeams.24.022802, DOWELL2003331, MALYUTIN2017105} are routinely used. 
For higher-dimensional structures, especially when correlations between the transverse and longitudinal phase spaces are present, the above mentioned techniques are insufficient. 
Tomographic methods have been successfully applied to these cases, e.g., for measuring the time-resolved transverse phase-space distribution  \cite{PhysRevSTAB.12.050704}, the full 4D transverse phase space \cite{HOCK20138, PhysRevAccelBeams.23.032804, guo:ibic2021-tupp15, jaster-merz:ipac21-mopab302}, as well as the full 3D spatial distribution \cite{Marx_2017, Marx2019, Marx:427780, Marchetti2021}.

The development of the polarizable X-band transverse-deflection structure (PolariX TDS) \cite{PhysRevAccelBeams.23.112001, Marchetti2021, Grudiev:2158484} by a collaboration of scientists from CERN, PSI, and DESY enables the extension of these existing diagnostics abilities \cite{Marx_2017, Marchetti2021, gonzalez_commissioning} to include streaking of the bunch in any transverse direction.
Based on this capability, we recently proposed a new method that extends the existing tomographic methods by an extra dimension and reconstructs the full 5D phase-space density of bunches \cite{Jaster-Merz_5D_article}, i.e., the transverse positions $(x, y)$ and divergences $(x', y')$ as well as the time coordinate $t$.
This technique combines a quadrupole-based 4D transverse tomography \cite{HOCK20138} in normalized phase space with the streaking of the bunch along various directions by the PolariX TDS.
Simulation studies of the 5D tomography method have shown the successful reconstruction of the 5D phase space of complex beam distributions \cite{Jaster-Merz_5D_article}. 

Other methods have been studied to obtain higher dimensional information. 
For sub-relativistic beams, a combination of horizontal and vertical slit masks can be used and has been benchmarked with a \SI{2.5}{MeV} $\mathrm{H}^{-}$ beam for 5D \cite{PhysRevAccelBeams.26.064202}, and 6D \cite{PhysRevLett.121.064804} reconstructions. 
Furthermore, machine learning based approaches are gaining popularity, which focus on predicting phase-space projections \cite{20230074, PhysRevLett.121.044801, Scheinker2022, PhysRevE.107.045302} up to the full 6D particle distribution \cite{PhysRevAccelBeams.27.094601}. 
While these methods show promising results, the obtained distribution is only a prediction ensured to match very limited beam measurements.

In this letter, we focus on the use of tomographic methods that are based on direct measurements of all relevant phase-space projections.
We present the first 5D tomographic reconstruction of an electron bunch as well as self-consistent checks to validate the results. 
We show that the method resolves previously hidden correlations between all the transverse planes and time coordinate and in addition allows the 4D slice emittance to be extracted. 
Furthermore, by propagating the reconstructed 5D distribution along the beamline, new possibilities open up for performing highly accurate simulation studies to understand and optimize the accelerator performance or to benchmark the results of simulation codes. 

\begin{figure*}
    \centering
    \includegraphics*[width=1.\textwidth]{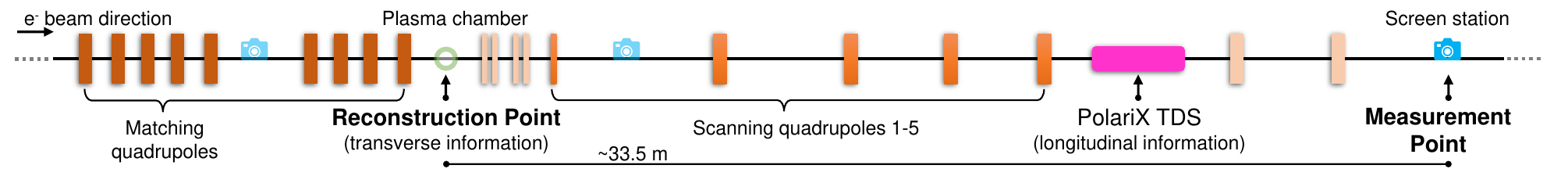}
    \caption{The FLASHForward beamline section used for the 5D tomography measurement. The transverse phase space is reconstructed at the location labeled as the reconstruction point. The reconstruction of the time information is obtained at the PolariX TDS. 
    The screen images for the tomography were recorded at the measurement point. Beamline elements that were not used for the measurement are displayed in a fainter color.}
    \label{fig:FF_beamline}
\end{figure*} 
The measurement campaign was performed at the FLASHForward facility \cite{d2019flashforward} which is dedicated to beam-driven plasma-acceleration experiments.
It uses the high-quality and stable electron bunches provided by the FLASH linear accelerator \cite{schreiber_faatz_2015, Ackermann2007, ROSSBACH20191}. 
A sketch of the relevant beamline section is shown in Fig.~\ref{fig:FF_beamline}.
To record the beam images, a screen station downstream of the PolariX TDS was used with Scheimpflug optics, a Cerium-doped Gadolinium Aluminum Gallium Garnet (GAGG:Ce) scintillating screen, and a nominal resolution of \SI[]{10}{\micro \meter} \cite{wiebers:ibic13-wepf03}.
Nine quadrupoles were operated to set the beam optics \cite{COURANT19581} to $\beta_{x} = \beta_{y} = \SI{10}{m}$ and $\alpha_{x} = \alpha_{y} = 0$ at the reconstruction point.
The 5D tomography method \cite{Jaster-Merz_5D_article} simultaneously controls the horizontal and vertical phase advances between the reconstruction point and measurement point.
These phase advances were varied using five quadrupoles covering a range of \SI{180}{\degree} in ten steps in both transverse planes. 
The required quadrupole strengths were determined in simulations. 
For each phase advance combination, the beam was streaked using the PolariX TDS at ten transverse angles covering approximately \SI{180}{\degree}.
This resulted in a total of 1000 planned scan steps. 
All quadrupoles between the TDS and measurement screen were switched off as identical beam optics in all transverse directions between these points are required for the tomographic reconstruction of the charge density \cite{Marx2019}. 

\begin{figure*}
    \centering
    \includegraphics*[width=1.0\textwidth]{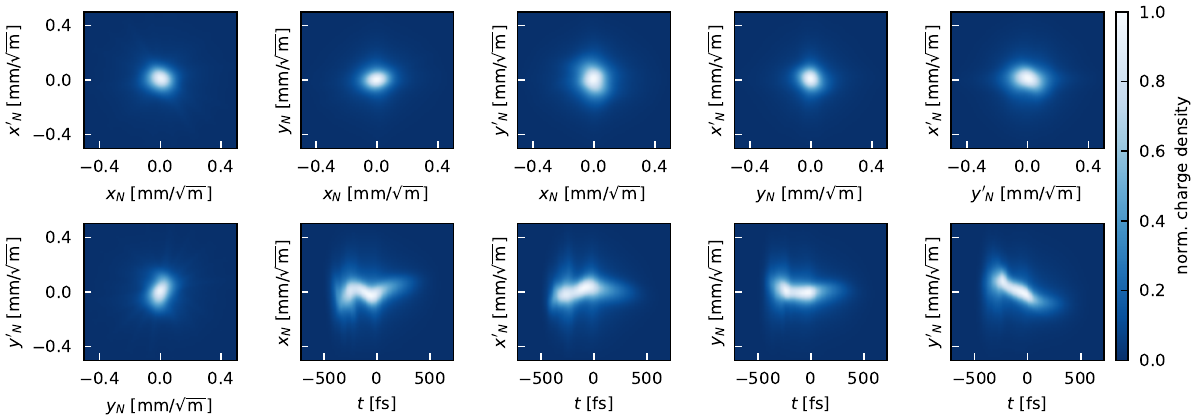}
    \caption{2D projections of the reconstructed 5D phase-space distribution normalized to their maximum value. The transverse projections are shown in normalized phase space. The head of the bunch is towards negative time values.}
    \label{fig:2D_projections}
\end{figure*}

\begin{figure}
    \centering
    \includegraphics*[width=1.\columnwidth]{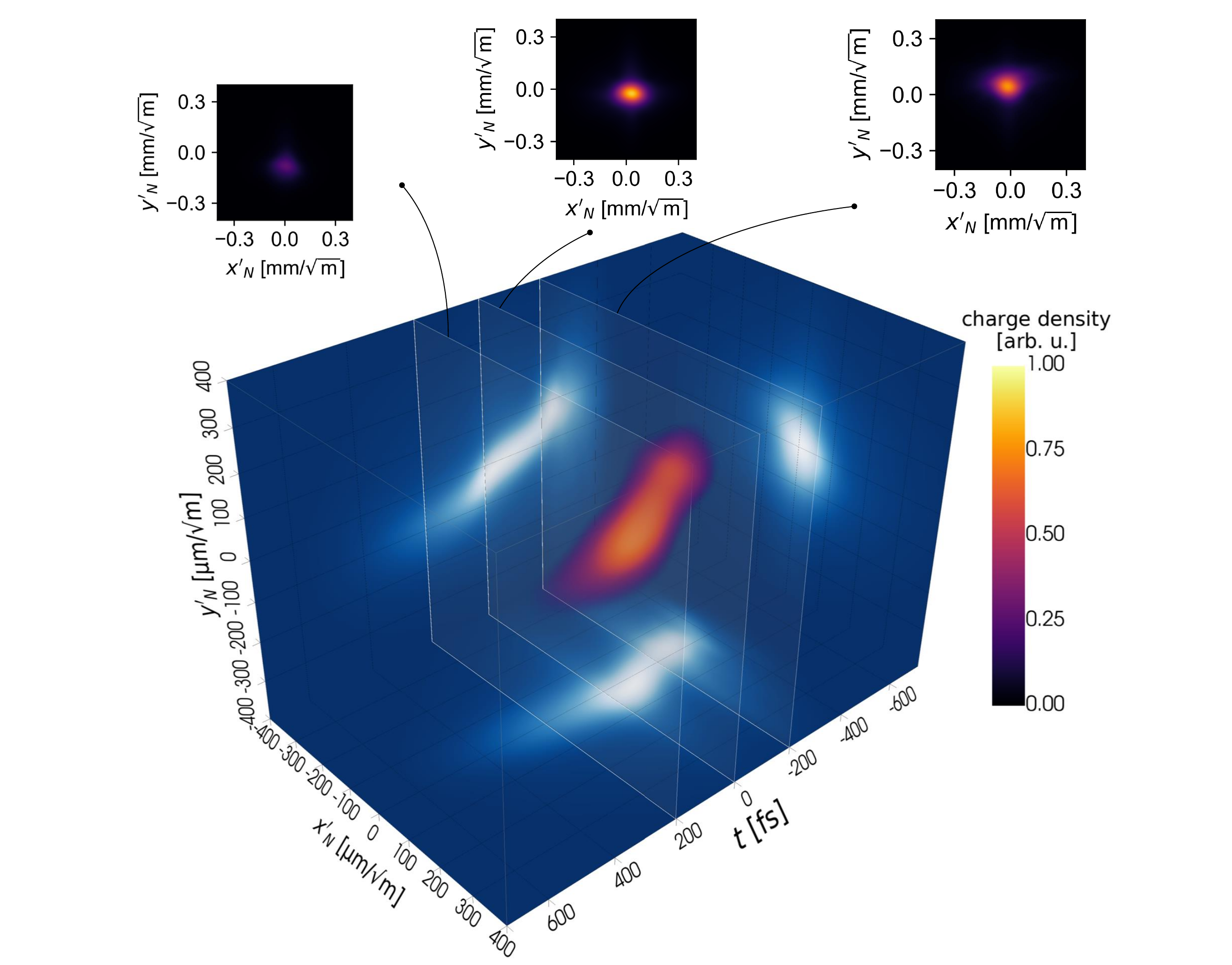}
    \caption{Rendering of the reconstructed 5D phase space distribution projected onto the $(x'_{N}, y'_{N} ,t$)
    phase space. The projection is normalized to the maximum value of the charge density. In addition, the 2D projections are displayed in blue color and exemplarily three transverse slices are included.}
    \label{fig:3D_render}
\end{figure}

The shear parameter, relating the measured transverse coordinate along the streaking direction at the measurement point to the longitudinal position within the bunch at the TDS location, was measured for all streaking angles by means of a radio-frequency (RF) phase scan.
The PolariX amplitude was set to the maximum of the available RF power at an average value of \SI{4.5}{\mega\watt} $\pm$ \SI{0.3}{\mega\watt} resulting in an average shear parameter of \SI{18.7}{} $\pm$ \SI{2.5}{}. 
The beta function at the measurement point was kept approximately at $\beta_{x} = \beta_{y} = \SI[]{10}{\meter}$ to obtain a near-constant longitudinal resolution in the tens-of-fs range while maintaining an achievable optics setup for the beamline.

For the 5D phase-space-tomography measurement, the quadrupoles were cycled between scan points whenever their ramping direction was changed to prevent hysteresis effects. 
For each scan point, ten screen images with beam as well as ten background images were recorded.
In addition, screen images of the unstreaked beam were collected.
During the experiment, 960 out of 1000 scan points were recorded successfully in \SI[]{28}{\hour}.
Charge and compression feedbacks were enabled to obtain a stable beam with RMS variations below \SI{1}{\percent} for the charge and below \SI{8}{\percent} for the bunch length over the entire measurement. 
The measurement was conducted at a beam energy of \SI{1.09}{\giga\electronvolt} $\pm$ \SI{0.01}{\giga\electronvolt} and with an average beam charge of \SI{297}{pC} $\pm$ \SI{2}{pC}. 

The 5D tomographic reconstruction procedure is described in detail in reference \cite{Jaster-Merz_5D_article}.
A Python scikit-image implementation of the SART (Simultaneous Algebraic Reconstruction Technique) algorithm \cite{ANDERSEN198481, van2014scikit} was used with two iterations.
In a first step, the 3D charge-density distribution was reconstructed \cite{Marx_2017} at the measurement location for each phase-advance combination.
This distribution was then converted into normalized coordinates using the Courant-Snyder parameters \cite{COURANT19581} at this location obtained from simulations. 
For each time slice of the bunch, the transverse distributions of all phase advance settings were combined and the 4D phase space was reconstructed using a method similar to the one presented in \cite{HOCK20138}.
Finally, by joining all time slices the 5D distribution $(x_{N}, x'_{N}, y_{N}, y'_{N}, t)$ was obtained.
The time information was reconstructed at the location of the PolariX TDS using 72 slices of \SI{20}{fs} duration, with the duration approximately corresponding to the average longitudinal resolution of \SI{19}{\femto\second} $\pm$ \SI{5}{\femto\second}. 
The transverse information was reconstructed upstream of the five scanning quadrupoles (see Fig.~\ref{fig:FF_beamline}) with 301 bins in each transverse direction in a range of $\pm$\SI[]{5e-4}{\meter\per\sqrt\meter} corresponding to an approximate bin size of the assumed \SI{10}{\micro\meter} screen resolution. 
A constant beam charge was assumed in all tomographic reconstructions.

The projection of the reconstructed 5D phase space onto the ten 2D planes is shown in Fig.~\ref{fig:2D_projections}, which constitutes the first direct and simultaneous measurements of these projections for ultra-relativistic electron beams. 
Non-linear correlations between all transverse planes and the time coordinate are visible. 
The projected normalized transverse phase spaces deviate from the matched circular case and exhibit non-Gaussian structures which are visible especially in the $(x'_{N}, y'_{N})$ and $(y_{N}, y'_{N})$ planes. 
These features mainly stem from varying centroid and momentum offsets along the bunch.
Such correlations can appear in linear accelerators and can be caused by collective effects, such as space charge and coherent synchrotron radiation (CSR) occurring, e.g., during the compression of the bunch \cite{Derbenev:414678, dohlus2006bunch}.
In particular, the correlations in the $(x_{N}, t)$ plane are expected to originate from CSR effects induced in the two horizontal bunch compressor chicanes located upstream in the FLASH linac, as qualitatively similar features have also been observed in previous simulation studies \cite{Marx:427780, Marx_2017}.
The correlations in the $(y'_{N}, t)$ phase space cannot be explained by these CSR effects.
We believe that these correlations are caused by the remaining vertical dispersion in the beamline which, in combination with the induced energy chirp during bunch compression, would lead to correlations between the vertical planes and the time coordinate. 
An optimization of the observed correlations was beyond the scope of this proof-of-principle demonstration but is now possible due to the ability of the 5D tomography method to unveil these effects. 

In addition to the 2D projections, the projection onto the 3D $(x'_{N}, y'_{N}, t)$ phase space is shown in Fig.~\ref{fig:3D_render}, highlighting three transverse slices along the bunch and the ability of the method to provide a time-resolved measurement of these less conventional transverse phase spaces.
The aforementioned features and non-linear correlations, as well as momentum offsets, are also visible here.

\begin{figure}
    \centering
    \includegraphics*[width=1.\columnwidth]{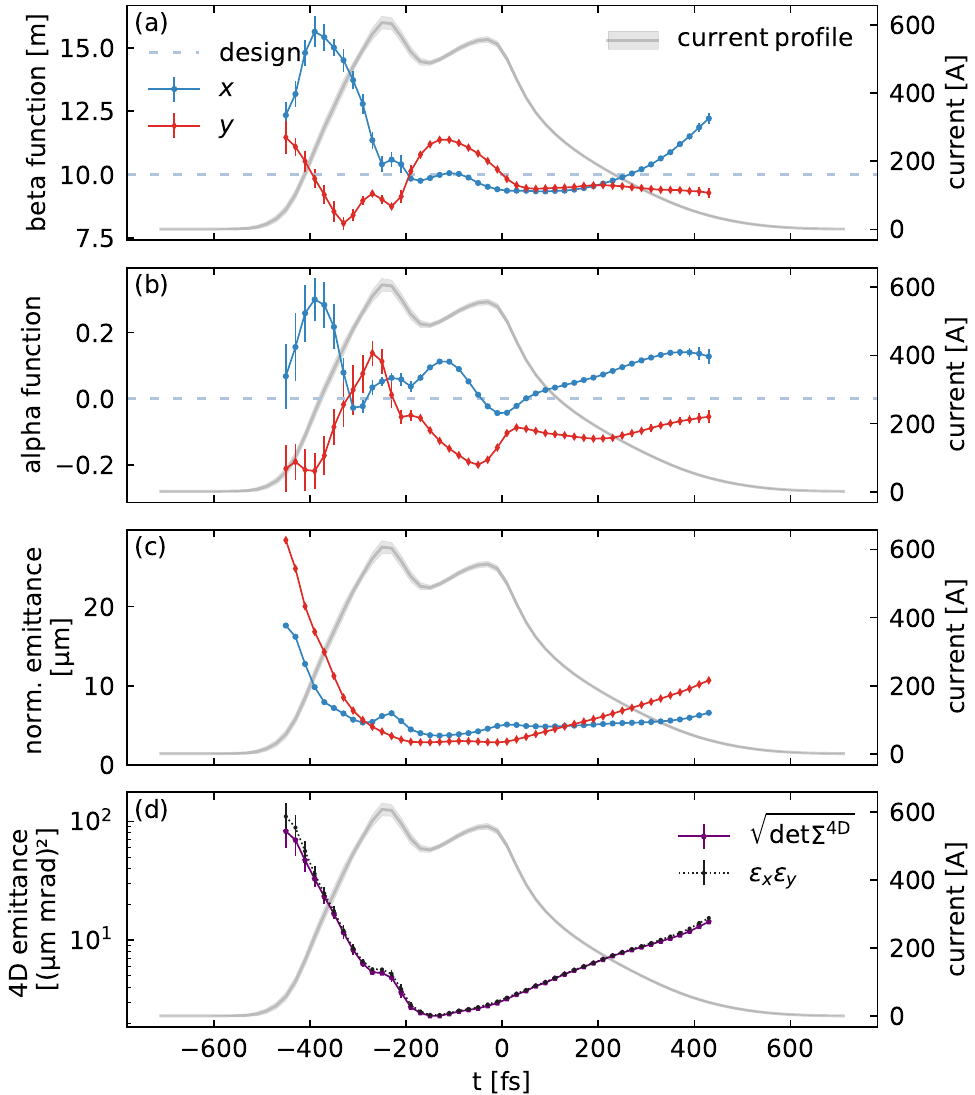}
    \caption{Reconstructed current profile, (a) beta functions, (b) alpha functions, and (c) sliced normalized transverse emittances for both transverse planes obtained from the 5D charge density. (d) The sliced 4D emittance is compared to the value obtained when multiplying the two transverse emittances. Only slices that contain at least \SI{10000}{particles} (\SI{0.2}{\percent} of the total charge) are analyzed and the errorbars are obtained from 100 reconstructions where the shear parameter of each streaking angle is randomly sampled from a Gaussian distribution with a one sigma measurement uncertainty.}
    \label{fig:sliced_beam_parameters}
\end{figure} 
Furthermore, the reconstruction can be used to create a particle distribution with a significant number of macro particles, in the presented case five million.
The distribution is generated slice-wise by randomly creating particles in accordance with the probability distribution of the reconstructed phase space.
Only probability values larger than \SI{5}{\percent} of the maximum value of each slice are appropriately populated.  
This threshold has been found to be a good balance between removing noise introduced by the tomographic reconstruction, typically occurring  far off-axis and arising due to the finite amount of projection angles \cite{Marchetti2021}, and preserving the features of the reconstructed distribution. 
The obtained distribution is used to calculate the beam covariance matrix $\Sigma^{4\mathrm{D}}$ for all slices containing at least \SI{10000}{particles} and extract the slice beam parameters shown in Fig.~\ref{fig:sliced_beam_parameters} like the transverse Courant-Snyder parameters, normalized emittances $\epsilon_{x}^{n}, \epsilon_{y}^{n}$ and the current profile. 
Variations of these parameters along the bunch are visible as well as a mismatch to the design values of $\beta = \SI{10}{\meter}$ and $\alpha = 0$. 
In addition, the sliced 4D emittance $\epsilon^{4\mathrm{D}} = \sqrt{\mathrm{det} \Sigma^{4\mathrm{D}}}$ is extracted. 
It is a measure of the minimum transverse emittances that can be reached in the case that all correlations between the planes are eliminated \cite{rivkin1986damping}. 
As the lasing performance of FELs and the luminosity of colliders depend on the transverse emittances \cite{PhysRevSTAB.17.052801, Aicheler:1500095}, having the ability to now measure and in the future correct for correlations is highly beneficial. 
The measured sliced 4D emittance is shown in Fig.~\ref{fig:sliced_beam_parameters}~(d) and compared to a multiplication of the two transverse emittances $\epsilon^{4\mathrm{D}}_{\mathrm{{approx.}}} = \epsilon_{x} \epsilon_{y}$, which are accessible with conventional PolariX TDS measurements. 
The average relative discrepancy is \SI{5}{\percent} with a minimum of \SI{1}{\percent} at around \SI{-170}{fs}.
The measurement therefore indicates that only minor correlations between the transverse planes are present which are minimal at the core part of the bunch. 

\begin{figure*}
    \centering
    \includegraphics*[width=1.\textwidth]{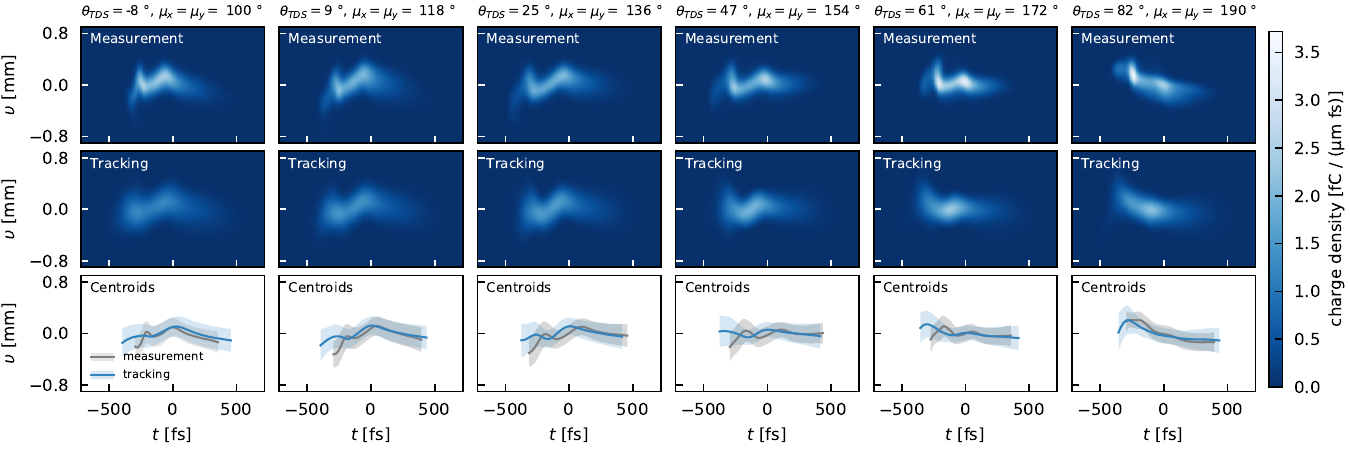}
    \caption{Comparison of the measured screen images (top) to the tracked distribution (middle) for various phase advances $\mu_{x}, \mu_{y}$. The screen images are rotated by the streaking angle $\theta_{TDS}$ and the projection in streaking direction is converted to the time axis. The coordinate $\upsilon$ denotes the transverse plane of the bunch perpendicular to the streaking plane. The bottom row shows the centroid positions and RMS spread for each time slice containing more than \SI{0.5}{\percent} of the total charge.}
    \label{fig:comparison_measurement_tracking}
\end{figure*} 
The tomographic reconstruction is validated using two approaches. 
First, the beam parameters obtained from the tomographic reconstruction are compared to lower dimensional complementary measurements listed in Table \ref{tab:beam_parameters}.
To obtain the normalized transverse RMS emittances and Courant-Snyder parameters, a multi-quadrupole scan analysis is performed using the unstreaked beam images.
The uncertainties are obtained from statistical fluctuation from ten different datasets recorded throughout the 5D tomography measurement.
The uncertainty on the tomographic reconstruction is estimated by performing 100 reconstructions using the Python library Optimas \cite{PhysRevAccelBeams.26.084601} where the shear parameter of each streaking angle is randomly sampled from a Gaussian distribution and a one sigma measurement uncertainty is applied. 
In addition, errors on the reconstruction due to an energy uncertainty (a \SI{1}{\percent} energy error results in a \SI{5}{\percent} emittance error), an energy spread within the distribution (a projected RMS energy spread of \SI{1}{\percent} consisting of a linear energy chirp and an uncorrelated RMS energy spread of \SI{0.1}{\percent} result in a \SI{2}{\percent} emittance error), and due to the finite number of projection angles (ten angles per plane result in an overestimation of the emittance by up to \SI{13}{\percent}) are expected. 
While these measurement errors are expected to apply equally to both transverse planes, we suspect that the larger discrepancy in the vertical emittance can be explained by an incomplete dataset of the last vertical phase advance scan (and hence a reduced number of recorded projections compared to the horizontal plane) resulting in an increased measurement uncertainty in this plane. 
The discrepancies in the alpha functions cannot be explained by the expected measurement uncertainties of the two methods. 
The beta functions and horizontal emittance agree well between the two methods within the estimated uncertainties. 
The bunch duration is determined using the screen images of the streaked beam at both zero crossings. 
Following \cite{emma:epac02-thpri097, Behrens2014}, linear correlations within the bunch are taken into account resulting in an RMS bunch duration of \SI[]{194}{fs} $\pm$ \SI[]{16}{fs} which is in excellent agreement with the result of $199^{+8}_{-7} \ \mathrm{fs}$ obtained from the tomography.

Second, the obtained particle distribution from the reconstructed 5D phase space is tracked from the reconstruction point to the measurement point for all beamline settings that were used in the measurement.
The obtained simulated screen images of the distribution are then compared to the measured ones.
This comparison is done in the $(\upsilon, t)$ coordinates, where $\upsilon$ is the transverse plane perpendicular to the streaking direction and $t$ is the time axis obtained by scaling the streaking plane with the shear parameter.
As a measure of deviation between the reconstructed and measured projections, the deviations between the centroid position of each time slice normalized to the RMS spread of the measured distribution are determined, and the average weighted by the slice charge is calculated.
Only slices that contain more than \SI{0.5}{\percent} of the total charge are taken into account.
The average relative discrepancy for all recorded projections is \SI{32}{\percent} $\pm$ \SI{15}{\percent}.
A visual comparison of the measured and tracked screen images shows that for most projections a qualitatively accurate reconstruction of the beam shape is obtained, as can be seen in a subset of the data shown in Fig.~\ref{fig:comparison_measurement_tracking}.
The main discrepancy between the measured and tracked projections is expected to stem from a blurring and loss of small beam features, which are especially sensitive to uncertainties in the shear parameter, as well as small changes in the distribution over the measurement time.

Overall, the two validations indicate a reasonable agreement of the reconstructed 5D distribution with lower-dimensional complementary measurements and thereby support the experimental usability of the 5D tomography method.

\begin{table}
    \begin{minipage}{\columnwidth}
	\centering
	\caption{Reconstructed Beam Parameters}
    \begin{ruledtabular}
		\begin{tabular}{ll c c} 
            \textbf{Param.} & \textbf{Units}  & \textbf{Tomography} & \textbf{Quad. scan} \\
            \colrule 
			$\epsilon_{x}^{n}$ & \si[]{\micro \metre} & 6.63 $\pm$ 0.10 & 6.22 $\pm$ 0.68\\
			$\epsilon_{y}^{n}$ & \si[]{\micro \metre} & 8.25 $\pm$ 0.20 & 6.69 $\pm$ 0.62\\ 
			$\alpha_{x}$ & - & 0.09 $\pm$ 0.00 & 0.37 $\pm$ 0.08 \\
			$\alpha_{y}$ & - & -0.28 $\pm$ 0.01 & 0.05 $\pm$ 0.14 \\
			$\beta_{x}$ & \si[]{\metre} & 10.75 $\pm$ 0.04 & 11.74 $\pm$ 1.50 \\
			$\beta_{y}$ & \si[]{\metre} & 7.44 $\pm$ 0.05 & 7.04 $\pm$ 0.61 \\
            \colrule
            $\sigma_{t}$ & \si{\femto\second} & $199^{+8}_{-7}$ & 194 $\pm$ 16 \footnote{Obtained from streaked screen images at both zero crossings.}\\
		\end{tabular}
	\label{tab:beam_parameters}
    \end{ruledtabular}
\end{minipage}
\end{table}

In this work we presented the first tomographic reconstruction of the 5D phase space of an electron beam and the validation of the result against lower dimensional projections. 
The 2D projections of the reconstructed phase space show correlations between the transverse and longitudinal planes that were not detectable with previous diagnostic methods.
Furthermore, the tomographic reconstruction facilitated the first sliced 4D emittance measurement at a free-electron laser. 
While the reconstructed bunch was not optimized for FEL or plasma operation, both applications can benefit from the presented measurement method as undesired correlations that either degrade the transverse emittances (and hence the lasing performance of FELs) or drive instabilities in beam-driven plasma accelerators \cite{PhysRevLett.67.991, PhysRevLett.99.255001, PhysRevLett.118.174801} can be identified and minimized.
Furthermore, conversion of the reconstructed phase space into a particle distribution enables highly realistic simulations of the beamline with the potential to better understand and improve its performance. 
A future combination of such a 5D tomography with an additional longitudinal phase space measurement would provide a quasi-complete characterization of electron beams.
These results can additionally be used to benchmark other diagnostic methods, e.g., based on machine learning \cite{PhysRevAccelBeams.27.094601} which use only minimal beam projections to infer the particle distribution.

To increase the viability of the measurement, a reduction of the measurement time is strongly desired.
During the measurement campaign we could already show that an optimized operation of the PolariX RF amplitude while changing the streaking angle reduces the measurement time by up to \SI{30}{\percent}. 
The remaining dominant factor was the cycling of the quadrupoles.
If the measurement is modified in such a way that no cycling is needed, the measurement time can be reduced substantially.
Proof-of-principle tests in this direction at the ARES accelerator \cite{burkart:linac22-thpojo01} have in fact been shown to reduce the measurement time by up to \SI{80}{\percent} \cite{jaster-merz:ipac21-mopab302}. 
Finally, the use of machine-learning-assisted reconstruction methods \cite{PhysRevResearch.6.033163, PhysRevAccelBeams.27.094601} or tomography algorithms based on particle sampling \cite{hoover2025ndimensionalmaximumentropytomographyparticle} could be explored to reduce the number of required projections. 

We would like to thank S. Schreiber and the scientific committee at FLASH for supporting this study and providing this beamtime, M. Vogt for the support on the PolariX structures, as well as the FLASH operators for their support during the beamtime. We thank the technical groups for their support, especially R. Jonas for helpful advice on the operation of the PolariX phase shifter. Furthermore, we thank A. Wolski for helpful discussions on optimizing the beamline for the tomographic measurement, A. Ferran Pousa for helpful discussions on the data analysis and the support on the optimas library, and B. Beutner for the helpful discussions on the results.
This research was supported in part through the Maxwell computational resources operated at DESY, Hamburg, Germany. We acknowledge support from DESY Hamburg, Germany, a member of the Helmholtz Association HGF.

SJM, RA, RB, FB, WH, MS, and TV conceptualized the work
and 
proposed the experiment. 
SJM, FB, PGC, and RD planned the experiment. 
SJM, PGC, FM, and SW conducted preparatory work for the experiment. 
SJM, JB, JBS, FB, PC, RD, HD, PGC, AK, MK, WK, FM, BS, TV, and SW conducted the experiment.
SJM performed the data analysis and wrote the manuscript.
JB, JBS, FB, PC, RD, HD, MK, WK, FM, TV, and SW contributed to the manuscript.
Everybody read the manuscript and contributed to discussions of the results. 

\bibliography{5D_measurement}

\begin{thebibliography}{66}%
\makeatletter
\providecommand \@ifxundefined [1]{%
 \@ifx{#1\undefined}
}%
\providecommand \@ifnum [1]{%
 \ifnum #1\expandafter \@firstoftwo
 \else \expandafter \@secondoftwo
 \fi
}%
\providecommand \@ifx [1]{%
 \ifx #1\expandafter \@firstoftwo
 \else \expandafter \@secondoftwo
 \fi
}%
\providecommand \natexlab [1]{#1}%
\providecommand \enquote  [1]{``#1''}%
\providecommand \bibnamefont  [1]{#1}%
\providecommand \bibfnamefont [1]{#1}%
\providecommand \citenamefont [1]{#1}%
\providecommand \href@noop [0]{\@secondoftwo}%
\providecommand \href [0]{\begingroup \@sanitize@url \@href}%
\providecommand \@href[1]{\@@startlink{#1}\@@href}%
\providecommand \@@href[1]{\endgroup#1\@@endlink}%
\providecommand \@sanitize@url [0]{\catcode `\\12\catcode `\$12\catcode
  `\&12\catcode `\#12\catcode `\^12\catcode `\_12\catcode `\%12\relax}%
\providecommand \@@startlink[1]{}%
\providecommand \@@endlink[0]{}%
\providecommand \url  [0]{\begingroup\@sanitize@url \@url }%
\providecommand \@url [1]{\endgroup\@href {#1}{\urlprefix }}%
\providecommand \urlprefix  [0]{URL }%
\providecommand \Eprint [0]{\href }%
\providecommand \doibase [0]{https://doi.org/}%
\providecommand \selectlanguage [0]{\@gobble}%
\providecommand \bibinfo  [0]{\@secondoftwo}%
\providecommand \bibfield  [0]{\@secondoftwo}%
\providecommand \translation [1]{[#1]}%
\providecommand \BibitemOpen [0]{}%
\providecommand \bibitemStop [0]{}%
\providecommand \bibitemNoStop [0]{.\EOS\space}%
\providecommand \EOS [0]{\spacefactor3000\relax}%
\providecommand \BibitemShut  [1]{\csname bibitem#1\endcsname}%
\let\auto@bib@innerbib\@empty
\bibitem [{\citenamefont {Chen}\ \emph {et~al.}(1985)\citenamefont {Chen} \emph
  {et~al.}}]{PhysRevLett.54.693}%
  \BibitemOpen
  \bibfield  {author} {\bibinfo {author} {\bibfnamefont {P.}~\bibnamefont
  {Chen}} \emph {et~al.},\ }\bibfield  {title} {\bibinfo {title} {Acceleration
  of electrons by the interaction of a bunched electron beam with a plasma},\
  }\href {https://doi.org/10.1103/PhysRevLett.54.693} {\bibfield  {journal}
  {\bibinfo  {journal} {Phys. Rev. Lett.}\ }\textbf {\bibinfo {volume} {54}},\
  \bibinfo {pages} {693} (\bibinfo {year} {1985})}\BibitemShut {NoStop}%
\bibitem [{\citenamefont {Whittum}\ \emph {et~al.}(1991)\citenamefont {Whittum}
  \emph {et~al.}}]{PhysRevLett.67.991}%
  \BibitemOpen
  \bibfield  {author} {\bibinfo {author} {\bibfnamefont {D.~H.}\ \bibnamefont
  {Whittum}} \emph {et~al.},\ }\bibfield  {title} {\bibinfo {title}
  {Electron-hose instability in the ion-focused regime},\ }\href
  {https://doi.org/10.1103/PhysRevLett.67.991} {\bibfield  {journal} {\bibinfo
  {journal} {Phys. Rev. Lett.}\ }\textbf {\bibinfo {volume} {67}},\ \bibinfo
  {pages} {991} (\bibinfo {year} {1991})}\BibitemShut {NoStop}%
\bibitem [{\citenamefont {Galletti}\ \emph {et~al.}(2022)\citenamefont
  {Galletti} \emph {et~al.}}]{PhysRevLett.129.234801}%
  \BibitemOpen
  \bibfield  {author} {\bibinfo {author} {\bibfnamefont {M.}~\bibnamefont
  {Galletti}} \emph {et~al.},\ }\bibfield  {title} {\bibinfo {title} {Stable
  operation of a free-electron laser driven by a plasma accelerator},\ }\href
  {https://doi.org/10.1103/PhysRevLett.129.234801} {\bibfield  {journal}
  {\bibinfo  {journal} {Phys. Rev. Lett.}\ }\textbf {\bibinfo {volume} {129}},\
  \bibinfo {pages} {234801} (\bibinfo {year} {2022})}\BibitemShut {NoStop}%
\bibitem [{\citenamefont {Joshi}\ \emph {et~al.}(2018)\citenamefont {Joshi}
  \emph {et~al.}}]{Joshi_2018}%
  \BibitemOpen
  \bibfield  {author} {\bibinfo {author} {\bibfnamefont {C.}~\bibnamefont
  {Joshi}} \emph {et~al.},\ }\bibfield  {title} {\bibinfo {title} {{Plasma
  wakefield acceleration experiments at FACET II}},\ }\href
  {https://doi.org/10.1088/1361-6587/aaa2e3} {\bibfield  {journal} {\bibinfo
  {journal} {Plasma Physics and Controlled Fusion}\ }\textbf {\bibinfo {volume}
  {60}},\ \bibinfo {pages} {034001} (\bibinfo {year} {2018})}\BibitemShut
  {NoStop}%
\bibitem [{\citenamefont {Adli}(2018)}]{Adli2018}%
  \BibitemOpen
  \bibfield  {author} {\bibinfo {author} {\bibfnamefont {E.~o.}\ \bibnamefont
  {Adli}},\ }\bibfield  {title} {\bibinfo {title} {Acceleration of electrons in
  the plasma wakefield of a proton bunch},\ }\href
  {https://doi.org/10.1038/s41586-018-0485-4} {\bibfield  {journal} {\bibinfo
  {journal} {Nature}\ }\textbf {\bibinfo {volume} {561}},\ \bibinfo {pages}
  {363} (\bibinfo {year} {2018})}\BibitemShut {NoStop}%
\bibitem [{\citenamefont {Blumenfeld}\ \emph {et~al.}(2007)\citenamefont
  {Blumenfeld} \emph {et~al.}}]{Blumenfeld2007}%
  \BibitemOpen
  \bibfield  {author} {\bibinfo {author} {\bibfnamefont {I.}~\bibnamefont
  {Blumenfeld}} \emph {et~al.},\ }\bibfield  {title} {\bibinfo {title} {Energy
  doubling of 42{\thinspace}{GeV} electrons in a metre-scale plasma wakefield
  accelerator},\ }\href {https://doi.org/10.1038/nature05538} {\bibfield
  {journal} {\bibinfo  {journal} {Nature}\ }\textbf {\bibinfo {volume} {445}},\
  \bibinfo {pages} {741} (\bibinfo {year} {2007})}\BibitemShut {NoStop}%
\bibitem [{\citenamefont {Litos}\ \emph {et~al.}(2014)\citenamefont {Litos}
  \emph {et~al.}}]{Litos2014}%
  \BibitemOpen
  \bibfield  {author} {\bibinfo {author} {\bibfnamefont {M.}~\bibnamefont
  {Litos}} \emph {et~al.},\ }\bibfield  {title} {\bibinfo {title}
  {High-efficiency acceleration of an electron beam in a plasma wakefield
  accelerator},\ }\href {https://doi.org/10.1038/nature13882} {\bibfield
  {journal} {\bibinfo  {journal} {Nature}\ }\textbf {\bibinfo {volume} {515}},\
  \bibinfo {pages} {92} (\bibinfo {year} {2014})}\BibitemShut {NoStop}%
\bibitem [{\citenamefont {Ferrario}\ \emph {et~al.}(2013)\citenamefont
  {Ferrario} \emph {et~al.}}]{FERRARIO2013183}%
  \BibitemOpen
  \bibfield  {author} {\bibinfo {author} {\bibfnamefont {M.}~\bibnamefont
  {Ferrario}} \emph {et~al.},\ }\bibfield  {title} {\bibinfo {title}
  {{SPARC\_LAB present and future}},\ }\href
  {https://doi.org/https://doi.org/10.1016/j.nimb.2013.03.049} {\bibfield
  {journal} {\bibinfo  {journal} {Nuclear Instruments and Methods in Physics
  Research Section B: Beam Interactions with Materials and Atoms}\ }\textbf
  {\bibinfo {volume} {309}},\ \bibinfo {pages} {183} (\bibinfo {year}
  {2013})}\BibitemShut {NoStop}%
\bibitem [{\citenamefont {Pompili}\ \emph {et~al.}(2022)\citenamefont {Pompili}
  \emph {et~al.}}]{Pompili2022}%
  \BibitemOpen
  \bibfield  {author} {\bibinfo {author} {\bibfnamefont {R.}~\bibnamefont
  {Pompili}} \emph {et~al.},\ }\bibfield  {title} {\bibinfo {title}
  {Free-electron lasing with compact beam-driven plasma wakefield
  accelerator},\ }\href {https://doi.org/10.1038/s41586-022-04589-1} {\bibfield
   {journal} {\bibinfo  {journal} {Nature}\ }\textbf {\bibinfo {volume}
  {605}},\ \bibinfo {pages} {659} (\bibinfo {year} {2022})}\BibitemShut
  {NoStop}%
\bibitem [{\citenamefont {Hidding}\ \emph {et~al.}(2023)\citenamefont {Hidding}
  \emph {et~al.}}]{photonics10020099}%
  \BibitemOpen
  \bibfield  {author} {\bibinfo {author} {\bibfnamefont {B.}~\bibnamefont
  {Hidding}} \emph {et~al.},\ }\bibfield  {title} {\bibinfo {title} {Progress
  in hybrid plasma wakefield acceleration},\ }\bibfield  {journal} {\bibinfo
  {journal} {Photonics}\ }\textbf {\bibinfo {volume} {10}},\ \href
  {https://doi.org/10.3390/photonics10020099} {10.3390/photonics10020099}
  (\bibinfo {year} {2023})\BibitemShut {NoStop}%
\bibitem [{\citenamefont {Rossbach}\ \emph {et~al.}(2019)\citenamefont
  {Rossbach} \emph {et~al.}}]{ROSSBACH20191}%
  \BibitemOpen
  \bibfield  {author} {\bibinfo {author} {\bibfnamefont {J.}~\bibnamefont
  {Rossbach}} \emph {et~al.},\ }\bibfield  {title} {\bibinfo {title} {{10 years
  of pioneering X-ray science at the Free-Electron Laser FLASH at DESY}},\
  }\href {https://doi.org/10.1016/j.physrep.2019.02.002} {\bibfield  {journal}
  {\bibinfo  {journal} {Phys. Rep.}\ }\textbf {\bibinfo {volume} {808}},\
  \bibinfo {pages} {1} (\bibinfo {year} {2019})}\BibitemShut {NoStop}%
\bibitem [{\citenamefont {Abela}\ \emph {et~al.}(2006)\citenamefont {Abela}
  \emph {et~al.}}]{Abela:77248}%
  \BibitemOpen
  \bibfield  {author} {\bibinfo {author} {\bibfnamefont {R.}~\bibnamefont
  {Abela}} \emph {et~al.},\ }\href {https://doi.org/10.3204/DESY_06-097} {\emph
  {\bibinfo {title} {{XFEL}: {T}he {E}uropean {X}-{R}ay {F}ree-{E}lectron
  {L}aser - {T}echnical {D}esign {R}eport}}}\ (\bibinfo  {publisher} {DESY},\
  \bibinfo {address} {Hamburg},\ \bibinfo {year} {2006})\ pp.\ \bibinfo {pages}
  {1--646}\BibitemShut {NoStop}%
\bibitem [{\citenamefont {Emma}\ \emph {et~al.}(2010)\citenamefont {Emma} \emph
  {et~al.}}]{Emma2010}%
  \BibitemOpen
  \bibfield  {author} {\bibinfo {author} {\bibfnamefont {P.}~\bibnamefont
  {Emma}} \emph {et~al.},\ }\bibfield  {title} {\bibinfo {title} {First lasing
  and operation of an {\aa}ngstrom-wavelength free-electron laser},\ }\href
  {https://doi.org/10.1038/nphoton.2010.176} {\bibfield  {journal} {\bibinfo
  {journal} {Nature Photonics}\ }\textbf {\bibinfo {volume} {4}},\ \bibinfo
  {pages} {641} (\bibinfo {year} {2010})}\BibitemShut {NoStop}%
\bibitem [{\citenamefont {Ishikawa}\ \emph {et~al.}(2012)\citenamefont
  {Ishikawa} \emph {et~al.}}]{Ishikawa2012}%
  \BibitemOpen
  \bibfield  {author} {\bibinfo {author} {\bibfnamefont {T.}~\bibnamefont
  {Ishikawa}} \emph {et~al.},\ }\bibfield  {title} {\bibinfo {title} {A compact
  x-ray free-electron laser emitting in the sub-{\aa}ngstr{\"o}m region},\
  }\href {https://doi.org/10.1038/nphoton.2012.141} {\bibfield  {journal}
  {\bibinfo  {journal} {Nature Photonics}\ }\textbf {\bibinfo {volume} {6}},\
  \bibinfo {pages} {540} (\bibinfo {year} {2012})}\BibitemShut {NoStop}%
\bibitem [{\citenamefont {Giannessi}\ \emph {et~al.}(2019)\citenamefont
  {Giannessi} \emph {et~al.}}]{giannessi:fel2019-thp079}%
  \BibitemOpen
  \bibfield  {author} {\bibinfo {author} {\bibfnamefont {L.}~\bibnamefont
  {Giannessi}} \emph {et~al.},\ }\bibfield  {title} {\bibinfo {title} {{Status
  and Perspectives of the FERMI FEL Facility (2019)}},\ }in\ \href
  {https://doi.org/10.18429/JACoW-FEL2019-THP079} {\emph {\bibinfo {booktitle}
  {Proc. FEL'19}}},\ \bibinfo {series and number} {\bibinfo {series} {Free
  Electron Laser Conference}\ No.~\bibinfo {number} {39}}\ (\bibinfo
  {publisher} {JACoW Publishing, Geneva, Switzerland},\ \bibinfo {year}
  {2019})\ pp.\ \bibinfo {pages} {742--745}\BibitemShut {NoStop}%
\bibitem [{\citenamefont {Kang}\ \emph {et~al.}(2017)\citenamefont {Kang} \emph
  {et~al.}}]{Kang2017}%
  \BibitemOpen
  \bibfield  {author} {\bibinfo {author} {\bibfnamefont {H.-S.}\ \bibnamefont
  {Kang}} \emph {et~al.},\ }\bibfield  {title} {\bibinfo {title} {Hard x-ray
  free-electron laser with femtosecond-scale timing jitter},\ }\href
  {https://doi.org/10.1038/s41566-017-0029-8} {\bibfield  {journal} {\bibinfo
  {journal} {Nature Photonics}\ }\textbf {\bibinfo {volume} {11}},\ \bibinfo
  {pages} {708} (\bibinfo {year} {2017})}\BibitemShut {NoStop}%
\bibitem [{\citenamefont {Prat}\ \emph {et~al.}(2020)\citenamefont {Prat} \emph
  {et~al.}}]{Prat2020}%
  \BibitemOpen
  \bibfield  {author} {\bibinfo {author} {\bibfnamefont {E.}~\bibnamefont
  {Prat}} \emph {et~al.},\ }\bibfield  {title} {\bibinfo {title} {A compact and
  cost-effective hard x-ray free-electron laser driven by a high-brightness and
  low-energy electron beam},\ }\href
  {https://doi.org/10.1038/s41566-020-00712-8} {\bibfield  {journal} {\bibinfo
  {journal} {Nature Photonics}\ }\textbf {\bibinfo {volume} {14}},\ \bibinfo
  {pages} {748} (\bibinfo {year} {2020})}\BibitemShut {NoStop}%
\bibitem [{\citenamefont {Liu}\ \emph {et~al.}(2022)\citenamefont {Liu} \emph
  {et~al.}}]{liu2021sxfel}%
  \BibitemOpen
  \bibfield  {author} {\bibinfo {author} {\bibfnamefont {B.}~\bibnamefont
  {Liu}} \emph {et~al.},\ }\bibfield  {title} {\bibinfo {title} {The sxfel
  upgrade: From test facility to user facility},\ }\bibfield  {journal}
  {\bibinfo  {journal} {Applied Sciences}\ }\textbf {\bibinfo {volume} {12}},\
  \href {https://doi.org/10.3390/app12010176} {10.3390/app12010176} (\bibinfo
  {year} {2022})\BibitemShut {NoStop}%
\bibitem [{\citenamefont {Ferrari}\ \emph {et~al.}(2016)\citenamefont {Ferrari}
  \emph {et~al.}}]{Ferrari2016}%
  \BibitemOpen
  \bibfield  {author} {\bibinfo {author} {\bibfnamefont {E.}~\bibnamefont
  {Ferrari}} \emph {et~al.},\ }\bibfield  {title} {\bibinfo {title} {Widely
  tunable two-colour seeded free-electron laser source for resonant-pump
  resonant-probe magnetic scattering},\ }\bibfield  {journal} {\bibinfo
  {journal} {Nature Communications}\ }\textbf {\bibinfo {volume} {7}},\ \href
  {https://doi.org/10.1038/ncomms10343} {10.1038/ncomms10343} (\bibinfo {year}
  {2016})\BibitemShut {NoStop}%
\bibitem [{\citenamefont {Prat}\ \emph {et~al.}(2022)\citenamefont {Prat} \emph
  {et~al.}}]{PhysRevResearch.4.L022025}%
  \BibitemOpen
  \bibfield  {author} {\bibinfo {author} {\bibfnamefont {E.}~\bibnamefont
  {Prat}} \emph {et~al.},\ }\bibfield  {title} {\bibinfo {title} {Widely
  tunable two-color x-ray free-electron laser pulses},\ }\href
  {https://doi.org/10.1103/PhysRevResearch.4.L022025} {\bibfield  {journal}
  {\bibinfo  {journal} {Phys. Rev. Res.}\ }\textbf {\bibinfo {volume} {4}},\
  \bibinfo {pages} {L022025} (\bibinfo {year} {2022})}\BibitemShut {NoStop}%
\bibitem [{\citenamefont {McKee}\ \emph {et~al.}(1995)\citenamefont {McKee}
  \emph {et~al.}}]{MCKEE1995264}%
  \BibitemOpen
  \bibfield  {author} {\bibinfo {author} {\bibfnamefont {C.}~\bibnamefont
  {McKee}} \emph {et~al.},\ }\bibfield  {title} {\bibinfo {title} {Phase space
  tomography of relativistic electron beams},\ }\href
  {https://doi.org/10.1016/0168-9002(94)01411-6} {\bibfield  {journal}
  {\bibinfo  {journal} {Nucl. Instrum. Methods Phys. Res., Sect. A}\ }\textbf
  {\bibinfo {volume} {358}},\ \bibinfo {pages} {264} (\bibinfo {year}
  {1995})}\BibitemShut {NoStop}%
\bibitem [{\citenamefont {Stratakis}\ \emph {et~al.}(2006)\citenamefont
  {Stratakis} \emph {et~al.}}]{PhysRevSTAB.9.112801}%
  \BibitemOpen
  \bibfield  {author} {\bibinfo {author} {\bibfnamefont {D.}~\bibnamefont
  {Stratakis}} \emph {et~al.},\ }\bibfield  {title} {\bibinfo {title}
  {Tomography as a diagnostic tool for phase space mapping of intense particle
  beams},\ }\href {https://doi.org/10.1103/PhysRevSTAB.9.112801} {\bibfield
  {journal} {\bibinfo  {journal} {Phys. Rev. Accel. Beams}\ }\textbf {\bibinfo
  {volume} {9}},\ \bibinfo {pages} {112801} (\bibinfo {year}
  {2006})}\BibitemShut {NoStop}%
\bibitem [{\citenamefont {Yakimenko}\ \emph {et~al.}(2003)\citenamefont
  {Yakimenko} \emph {et~al.}}]{PhysRevSTAB.6.122801}%
  \BibitemOpen
  \bibfield  {author} {\bibinfo {author} {\bibfnamefont {V.}~\bibnamefont
  {Yakimenko}} \emph {et~al.},\ }\bibfield  {title} {\bibinfo {title} {Electron
  beam phase-space measurement using a high-precision tomography technique},\
  }\href {https://doi.org/10.1103/PhysRevSTAB.6.122801} {\bibfield  {journal}
  {\bibinfo  {journal} {Phys. Rev. Accel. Beams}\ }\textbf {\bibinfo {volume}
  {6}},\ \bibinfo {pages} {122801} (\bibinfo {year} {2003})}\BibitemShut
  {NoStop}%
\bibitem [{\citenamefont {Hermann}\ \emph {et~al.}(2021)\citenamefont {Hermann}
  \emph {et~al.}}]{PhysRevAccelBeams.24.022802}%
  \BibitemOpen
  \bibfield  {author} {\bibinfo {author} {\bibfnamefont {B.}~\bibnamefont
  {Hermann}} \emph {et~al.},\ }\bibfield  {title} {\bibinfo {title} {Electron
  beam transverse phase space tomography using nanofabricated wire scanners
  with submicrometer resolution},\ }\href
  {https://doi.org/10.1103/PhysRevAccelBeams.24.022802} {\bibfield  {journal}
  {\bibinfo  {journal} {Phys. Rev. Accel. Beams}\ }\textbf {\bibinfo {volume}
  {24}},\ \bibinfo {pages} {022802} (\bibinfo {year} {2021})}\BibitemShut
  {NoStop}%
\bibitem [{\citenamefont {Dowell}\ \emph {et~al.}(2003)\citenamefont {Dowell}
  \emph {et~al.}}]{DOWELL2003331}%
  \BibitemOpen
  \bibfield  {author} {\bibinfo {author} {\bibfnamefont {D.}~\bibnamefont
  {Dowell}} \emph {et~al.},\ }\bibfield  {title} {\bibinfo {title}
  {{Longitudinal emittance measurements at the SLAC gun test facility}},\
  }\href {https://doi.org/10.1016/S0168-9002(03)00940-9} {\bibfield  {journal}
  {\bibinfo  {journal} {Nucl. Instrum. Methods Phys. Res., Sect. A}\ }\textbf
  {\bibinfo {volume} {507}},\ \bibinfo {pages} {331} (\bibinfo {year}
  {2003})},\ \bibinfo {note} {proceedings of the 24th International Free
  Electron Laser Conference and the 9th Users Workshop.}\BibitemShut {Stop}%
\bibitem [{\citenamefont {Malyutin}\ \emph {et~al.}(2017)\citenamefont
  {Malyutin} \emph {et~al.}}]{MALYUTIN2017105}%
  \BibitemOpen
  \bibfield  {author} {\bibinfo {author} {\bibfnamefont {D.}~\bibnamefont
  {Malyutin}} \emph {et~al.},\ }\bibfield  {title} {\bibinfo {title}
  {{Longitudinal phase space tomography using a booster cavity at PITZ}},\
  }\href {https://doi.org/10.1016/j.nima.2017.07.043} {\bibfield  {journal}
  {\bibinfo  {journal} {Nucl. Instrum. Methods Phys. Res., Sect. A}\ }\textbf
  {\bibinfo {volume} {871}},\ \bibinfo {pages} {105} (\bibinfo {year}
  {2017})}\BibitemShut {NoStop}%
\bibitem [{\citenamefont {R\"ohrs}\ \emph {et~al.}(2009)\citenamefont
  {R\"ohrs}, \citenamefont {Gerth}, \citenamefont {Schlarb}, \citenamefont
  {Schmidt},\ and\ \citenamefont {Schm\"user}}]{PhysRevSTAB.12.050704}%
  \BibitemOpen
  \bibfield  {author} {\bibinfo {author} {\bibfnamefont {M.}~\bibnamefont
  {R\"ohrs}}, \bibinfo {author} {\bibfnamefont {C.}~\bibnamefont {Gerth}},
  \bibinfo {author} {\bibfnamefont {H.}~\bibnamefont {Schlarb}}, \bibinfo
  {author} {\bibfnamefont {B.}~\bibnamefont {Schmidt}},\ and\ \bibinfo {author}
  {\bibfnamefont {P.}~\bibnamefont {Schm\"user}},\ }\bibfield  {title}
  {\bibinfo {title} {Time-resolved electron beam phase space tomography at a
  soft x-ray free-electron laser},\ }\href
  {https://doi.org/10.1103/PhysRevSTAB.12.050704} {\bibfield  {journal}
  {\bibinfo  {journal} {Phys. Rev. Accel. Beams}\ }\textbf {\bibinfo {volume}
  {12}},\ \bibinfo {pages} {050704} (\bibinfo {year} {2009})}\BibitemShut
  {NoStop}%
\bibitem [{\citenamefont {Hock}\ \emph {et~al.}(2013)\citenamefont {Hock} \emph
  {et~al.}}]{HOCK20138}%
  \BibitemOpen
  \bibfield  {author} {\bibinfo {author} {\bibfnamefont {K.}~\bibnamefont
  {Hock}} \emph {et~al.},\ }\bibfield  {title} {\bibinfo {title} {Tomographic
  reconstruction of the full 4{D} transverse phase space},\ }\href
  {https://doi.org/10.1016/j.nima.2013.05.004} {\bibfield  {journal} {\bibinfo
  {journal} {Nucl. Instrum. Methods Phys. Res., Sect. A}\ }\textbf {\bibinfo
  {volume} {726}},\ \bibinfo {pages} {8} (\bibinfo {year} {2013})}\BibitemShut
  {NoStop}%
\bibitem [{\citenamefont {Wolski}\ \emph {et~al.}(2020)\citenamefont {Wolski}
  \emph {et~al.}}]{PhysRevAccelBeams.23.032804}%
  \BibitemOpen
  \bibfield  {author} {\bibinfo {author} {\bibfnamefont {A.}~\bibnamefont
  {Wolski}} \emph {et~al.},\ }\bibfield  {title} {\bibinfo {title} {Transverse
  phase space characterization in an accelerator test facility},\ }\href
  {https://doi.org/10.1103/PhysRevAccelBeams.23.032804} {\bibfield  {journal}
  {\bibinfo  {journal} {Phys. Rev. Accel. Beams}\ }\textbf {\bibinfo {volume}
  {23}},\ \bibinfo {pages} {032804} (\bibinfo {year} {2020})}\BibitemShut
  {NoStop}%
\bibitem [{\citenamefont {Guo}\ \emph {et~al.}(2021)\citenamefont {Guo},
  \citenamefont {Denham}, \citenamefont {Musumeci}, \citenamefont {Ody},\ and\
  \citenamefont {Park}}]{guo:ibic2021-tupp15}%
  \BibitemOpen
  \bibfield  {author} {\bibinfo {author} {\bibfnamefont {V.}~\bibnamefont
  {Guo}}, \bibinfo {author} {\bibfnamefont {P.}~\bibnamefont {Denham}},
  \bibinfo {author} {\bibfnamefont {P.}~\bibnamefont {Musumeci}}, \bibinfo
  {author} {\bibfnamefont {A.}~\bibnamefont {Ody}},\ and\ \bibinfo {author}
  {\bibfnamefont {Y.}~\bibnamefont {Park}},\ }\bibfield  {title} {\bibinfo
  {title} {{4D Beam Tomography at the UCLA Pegasus Laboratory}},\ }in\ \href
  {https://doi.org/10.18429/JACoW-IBIC2021-TUPP15} {\emph {\bibinfo {booktitle}
  {Proc. IBIC'21}}},\ \bibinfo {series and number} {International Beam
  Instrumentation Conference}\ (\bibinfo  {publisher} {JACoW Publishing,
  Geneva, Switzerland},\ \bibinfo {year} {2021})\ pp.\ \bibinfo {pages}
  {227--231}\BibitemShut {NoStop}%
\bibitem [{\citenamefont {Jaster-Merz}\ \emph {et~al.}(2021)\citenamefont
  {Jaster-Merz} \emph {et~al.}}]{jaster-merz:ipac21-mopab302}%
  \BibitemOpen
  \bibfield  {author} {\bibinfo {author} {\bibfnamefont {S.}~\bibnamefont
  {Jaster-Merz}} \emph {et~al.},\ }\bibfield  {title} {\bibinfo {title}
  {Characterization of the full transverse phase space of electron bunches at
  {ARES}},\ }in\ \href {https://doi.org/10.18429/JACoW-IPAC2021-MOPAB302}
  {\emph {\bibinfo {booktitle} {Proc. IPAC'21}}},\ \bibinfo {series and number}
  {12th Int. Particle Accelerator Conf.(IPAC’21), Campinas, Brazil}\
  (\bibinfo {year} {2021})\BibitemShut {NoStop}%
\bibitem [{\citenamefont {Marx}\ \emph {et~al.}(2017)\citenamefont {Marx} \emph
  {et~al.}}]{Marx_2017}%
  \BibitemOpen
  \bibfield  {author} {\bibinfo {author} {\bibfnamefont {D.}~\bibnamefont
  {Marx}} \emph {et~al.},\ }\bibfield  {title} {\bibinfo {title}
  {Reconstruction of the 3{D} charge distribution of an electron bunch using a
  novel variable-polarization transverse deflecting structure ({TDS})},\ }\href
  {https://doi.org/10.1088/1742-6596/874/1/012077} {\bibfield  {journal}
  {\bibinfo  {journal} {J. Phys. Conf. Ser.}\ }\textbf {\bibinfo {volume}
  {874}},\ \bibinfo {pages} {012077} (\bibinfo {year} {2017})}\BibitemShut
  {NoStop}%
\bibitem [{\citenamefont {Marx}\ \emph {et~al.}(2019)\citenamefont {Marx} \emph
  {et~al.}}]{Marx2019}%
  \BibitemOpen
  \bibfield  {author} {\bibinfo {author} {\bibfnamefont {D.}~\bibnamefont
  {Marx}} \emph {et~al.},\ }\bibfield  {title} {\bibinfo {title} {Simulation
  studies for characterizing ultrashort bunches using novel polarizable
  {X}-band transverse deflection structures},\ }\href
  {https://doi.org/10.1038/s41598-019-56433-8} {\bibfield  {journal} {\bibinfo
  {journal} {Sci. Rep.}\ }\textbf {\bibinfo {volume} {9}},\ \bibinfo {pages}
  {19912} (\bibinfo {year} {2019})}\BibitemShut {NoStop}%
\bibitem [{\citenamefont {Marx}(2019)}]{Marx:427780}%
  \BibitemOpen
  \bibfield  {author} {\bibinfo {author} {\bibfnamefont {D.}~\bibnamefont
  {Marx}},\ }\emph {\bibinfo {title} {{C}haracterization of {U}ltrashort
  {E}lectron {B}unches at the {SINBAD}-{ARES} {L}inac}},\ \href
  {https://doi.org/10.3204/PUBDB-2019-04190} {\bibinfo {type} {Dissertation}},\
  \bibinfo  {school} {Universit{\"a}t Hamburg}, \bibinfo {address} {Hamburg,
  Germany} (\bibinfo {year} {2019})\BibitemShut {NoStop}%
\bibitem [{\citenamefont {Marchetti}\ \emph {et~al.}(2021)\citenamefont
  {Marchetti} \emph {et~al.}}]{Marchetti2021}%
  \BibitemOpen
  \bibfield  {author} {\bibinfo {author} {\bibfnamefont {B.}~\bibnamefont
  {Marchetti}} \emph {et~al.},\ }\bibfield  {title} {\bibinfo {title}
  {Experimental demonstration of novel beam characterization using a
  polarizable {X}-band transverse deflection structure},\ }\href
  {https://doi.org/10.1038/s41598-021-82687-2} {\bibfield  {journal} {\bibinfo
  {journal} {Sci. Rep.}\ }\textbf {\bibinfo {volume} {11}},\ \bibinfo {pages}
  {3560} (\bibinfo {year} {2021})}\BibitemShut {NoStop}%
\bibitem [{\citenamefont {Craievich}\ \emph {et~al.}(2020)\citenamefont
  {Craievich} \emph {et~al.}}]{PhysRevAccelBeams.23.112001}%
  \BibitemOpen
  \bibfield  {author} {\bibinfo {author} {\bibfnamefont {P.}~\bibnamefont
  {Craievich}} \emph {et~al.},\ }\bibfield  {title} {\bibinfo {title} {Novel
  {$X$}-band transverse deflection structure with variable polarization},\
  }\href {https://doi.org/10.1103/PhysRevAccelBeams.23.112001} {\bibfield
  {journal} {\bibinfo  {journal} {Phys. Rev. Accel. Beams}\ }\textbf {\bibinfo
  {volume} {23}},\ \bibinfo {pages} {112001} (\bibinfo {year}
  {2020})}\BibitemShut {NoStop}%
\bibitem [{\citenamefont {Grudiev}(2016)}]{Grudiev:2158484}%
  \BibitemOpen
  \bibfield  {author} {\bibinfo {author} {\bibfnamefont {A.}~\bibnamefont
  {Grudiev}},\ }\bibfield  {title} {\bibinfo {title} {{design of compact high
  power rf components at x-band}},\ }\href {https://cds.cern.ch/record/2158484}
  {\bibfield  {journal} {\bibinfo  {journal} {CLIC note}\ } (\bibinfo {year}
  {2016})}\BibitemShut {NoStop}%
\bibitem [{\citenamefont {Gonz\'alez~Caminal}\ \emph
  {et~al.}(2024)\citenamefont {Gonz\'alez~Caminal} \emph
  {et~al.}}]{gonzalez_commissioning}%
  \BibitemOpen
  \bibfield  {author} {\bibinfo {author} {\bibfnamefont {P.}~\bibnamefont
  {Gonz\'alez~Caminal}} \emph {et~al.},\ }\bibfield  {title} {\bibinfo {title}
  {Beam-based commissioning of a novel $x$-band transverse deflection structure
  with variable polarization},\ }\href
  {https://doi.org/10.1103/PhysRevAccelBeams.27.032801} {\bibfield  {journal}
  {\bibinfo  {journal} {Phys. Rev. Accel. Beams}\ }\textbf {\bibinfo {volume}
  {27}},\ \bibinfo {pages} {032801} (\bibinfo {year} {2024})}\BibitemShut
  {NoStop}%
\bibitem [{\citenamefont {Jaster-Merz}\ \emph {et~al.}(2024)\citenamefont
  {Jaster-Merz}, \citenamefont {Assmann}, \citenamefont {Brinkmann},
  \citenamefont {Burkart}, \citenamefont {Hillert}, \citenamefont {Stanitzki},\
  and\ \citenamefont {Vinatier}}]{Jaster-Merz_5D_article}%
  \BibitemOpen
  \bibfield  {author} {\bibinfo {author} {\bibfnamefont {S.}~\bibnamefont
  {Jaster-Merz}}, \bibinfo {author} {\bibfnamefont {R.~W.}\ \bibnamefont
  {Assmann}}, \bibinfo {author} {\bibfnamefont {R.}~\bibnamefont {Brinkmann}},
  \bibinfo {author} {\bibfnamefont {F.}~\bibnamefont {Burkart}}, \bibinfo
  {author} {\bibfnamefont {W.}~\bibnamefont {Hillert}}, \bibinfo {author}
  {\bibfnamefont {M.}~\bibnamefont {Stanitzki}},\ and\ \bibinfo {author}
  {\bibfnamefont {T.}~\bibnamefont {Vinatier}},\ }\bibfield  {title} {\bibinfo
  {title} {{5D tomographic phase-space reconstruction of particle bunches}},\
  }\href {https://doi.org/10.1103/PhysRevAccelBeams.27.072801} {\bibfield
  {journal} {\bibinfo  {journal} {Phys. Rev. Accel. Beams}\ }\textbf {\bibinfo
  {volume} {27}},\ \bibinfo {pages} {072801} (\bibinfo {year}
  {2024})}\BibitemShut {NoStop}%
\bibitem [{\citenamefont {Hoover}\ \emph {et~al.}(2023)\citenamefont {Hoover},
  \citenamefont {Ruisard}, \citenamefont {Aleksandrov}, \citenamefont
  {Zhukov},\ and\ \citenamefont {Cousineau}}]{PhysRevAccelBeams.26.064202}%
  \BibitemOpen
  \bibfield  {author} {\bibinfo {author} {\bibfnamefont {A.}~\bibnamefont
  {Hoover}}, \bibinfo {author} {\bibfnamefont {K.}~\bibnamefont {Ruisard}},
  \bibinfo {author} {\bibfnamefont {A.}~\bibnamefont {Aleksandrov}}, \bibinfo
  {author} {\bibfnamefont {A.}~\bibnamefont {Zhukov}},\ and\ \bibinfo {author}
  {\bibfnamefont {S.}~\bibnamefont {Cousineau}},\ }\bibfield  {title} {\bibinfo
  {title} {Analysis of a hadron beam in five-dimensional phase space},\ }\href
  {https://doi.org/10.1103/PhysRevAccelBeams.26.064202} {\bibfield  {journal}
  {\bibinfo  {journal} {Phys. Rev. Accel. Beams}\ }\textbf {\bibinfo {volume}
  {26}},\ \bibinfo {pages} {064202} (\bibinfo {year} {2023})}\BibitemShut
  {NoStop}%
\bibitem [{\citenamefont {Cathey}\ \emph {et~al.}(2018)\citenamefont {Cathey},
  \citenamefont {Cousineau}, \citenamefont {Aleksandrov},\ and\ \citenamefont
  {Zhukov}}]{PhysRevLett.121.064804}%
  \BibitemOpen
  \bibfield  {author} {\bibinfo {author} {\bibfnamefont {B.}~\bibnamefont
  {Cathey}}, \bibinfo {author} {\bibfnamefont {S.}~\bibnamefont {Cousineau}},
  \bibinfo {author} {\bibfnamefont {A.}~\bibnamefont {Aleksandrov}},\ and\
  \bibinfo {author} {\bibfnamefont {A.}~\bibnamefont {Zhukov}},\ }\bibfield
  {title} {\bibinfo {title} {First six dimensional phase space measurement of
  an accelerator beam},\ }\href
  {https://doi.org/10.1103/PhysRevLett.121.064804} {\bibfield  {journal}
  {\bibinfo  {journal} {Phys. Rev. Lett.}\ }\textbf {\bibinfo {volume} {121}},\
  \bibinfo {pages} {064804} (\bibinfo {year} {2018})}\BibitemShut {NoStop}%
\bibitem [{\citenamefont {Yutao}\ \emph {et~al.}(2023)\citenamefont {Yutao},
  \citenamefont {Renkai},\ and\ \citenamefont {Weishi}}]{20230074}%
  \BibitemOpen
  \bibfield  {author} {\bibinfo {author} {\bibfnamefont {H.}~\bibnamefont
  {Yutao}}, \bibinfo {author} {\bibfnamefont {L.}~\bibnamefont {Renkai}},\ and\
  \bibinfo {author} {\bibfnamefont {W.}~\bibnamefont {Weishi}},\ }\href
  {https://doi.org/10.11884/HPLPB202335.230074} {\bibinfo {title} {Measurement
  of transverse phase space based on machine learning}} (\bibinfo {year}
  {2023})\BibitemShut {NoStop}%
\bibitem [{\citenamefont {Scheinker}\ \emph {et~al.}(2018)\citenamefont
  {Scheinker}, \citenamefont {Edelen}, \citenamefont {Bohler}, \citenamefont
  {Emma},\ and\ \citenamefont {Lutman}}]{PhysRevLett.121.044801}%
  \BibitemOpen
  \bibfield  {author} {\bibinfo {author} {\bibfnamefont {A.}~\bibnamefont
  {Scheinker}}, \bibinfo {author} {\bibfnamefont {A.}~\bibnamefont {Edelen}},
  \bibinfo {author} {\bibfnamefont {D.}~\bibnamefont {Bohler}}, \bibinfo
  {author} {\bibfnamefont {C.}~\bibnamefont {Emma}},\ and\ \bibinfo {author}
  {\bibfnamefont {A.}~\bibnamefont {Lutman}},\ }\bibfield  {title} {\bibinfo
  {title} {Demonstration of model-independent control of the longitudinal phase
  space of electron beams in the linac-coherent light source with femtosecond
  resolution},\ }\href {https://doi.org/10.1103/PhysRevLett.121.044801}
  {\bibfield  {journal} {\bibinfo  {journal} {Phys. Rev. Lett.}\ }\textbf
  {\bibinfo {volume} {121}},\ \bibinfo {pages} {044801} (\bibinfo {year}
  {2018})}\BibitemShut {NoStop}%
\bibitem [{\citenamefont {Scheinker}\ \emph {et~al.}(2022)\citenamefont
  {Scheinker}, \citenamefont {Cropp~V},\ and\ \citenamefont
  {Filippetto}}]{Scheinker2022}%
  \BibitemOpen
  \bibfield  {author} {\bibinfo {author} {\bibfnamefont {A.}~\bibnamefont
  {Scheinker}}, \bibinfo {author} {\bibfnamefont {F.~W.}\ \bibnamefont
  {Cropp~V}},\ and\ \bibinfo {author} {\bibfnamefont {D.}~\bibnamefont
  {Filippetto}},\ }\bibfield  {title} {\bibinfo {title} {{6D Phase Space
  Diagnostics Based on Adaptively Tuned Physics-Informed Generative
  Convolutional Neural Networks}},\ }in\ \href
  {https://doi.org/10.18429/JACoW-IPAC2022-TUOXGD3} {\emph {\bibinfo
  {booktitle} {Proc. IPAC'22}}},\ \bibinfo {series and number} {Proceedings of
  the 13th International Particle Accelerator Conference, Bangkok, Thailand}\
  (\bibinfo {year} {2022})\ pp.\ \bibinfo {pages} {776--779}\BibitemShut
  {NoStop}%
\bibitem [{\citenamefont {Scheinker}\ \emph {et~al.}(2023)\citenamefont
  {Scheinker}, \citenamefont {Cropp},\ and\ \citenamefont
  {Filippetto}}]{PhysRevE.107.045302}%
  \BibitemOpen
  \bibfield  {author} {\bibinfo {author} {\bibfnamefont {A.}~\bibnamefont
  {Scheinker}}, \bibinfo {author} {\bibfnamefont {F.}~\bibnamefont {Cropp}},\
  and\ \bibinfo {author} {\bibfnamefont {D.}~\bibnamefont {Filippetto}},\
  }\bibfield  {title} {\bibinfo {title} {Adaptive autoencoder latent space
  tuning for more robust machine learning beyond the training set for
  six-dimensional phase space diagnostics of a time-varying ultrafast
  electron-diffraction compact accelerator},\ }\href
  {https://doi.org/10.1103/PhysRevE.107.045302} {\bibfield  {journal} {\bibinfo
   {journal} {Phys. Rev. E}\ }\textbf {\bibinfo {volume} {107}},\ \bibinfo
  {pages} {045302} (\bibinfo {year} {2023})}\BibitemShut {NoStop}%
\bibitem [{\citenamefont {Roussel}\ \emph {et~al.}(2024)\citenamefont
  {Roussel}, \citenamefont {Gonzalez-Aguilera}, \citenamefont {Wisniewski},
  \citenamefont {Ody}, \citenamefont {Liu}, \citenamefont {Power},
  \citenamefont {Kim},\ and\ \citenamefont
  {Edelen}}]{PhysRevAccelBeams.27.094601}%
  \BibitemOpen
  \bibfield  {author} {\bibinfo {author} {\bibfnamefont {R.}~\bibnamefont
  {Roussel}}, \bibinfo {author} {\bibfnamefont {J.~P.}\ \bibnamefont
  {Gonzalez-Aguilera}}, \bibinfo {author} {\bibfnamefont {E.}~\bibnamefont
  {Wisniewski}}, \bibinfo {author} {\bibfnamefont {A.}~\bibnamefont {Ody}},
  \bibinfo {author} {\bibfnamefont {W.}~\bibnamefont {Liu}}, \bibinfo {author}
  {\bibfnamefont {J.}~\bibnamefont {Power}}, \bibinfo {author} {\bibfnamefont
  {Y.-K.}\ \bibnamefont {Kim}},\ and\ \bibinfo {author} {\bibfnamefont
  {A.}~\bibnamefont {Edelen}},\ }\bibfield  {title} {\bibinfo {title}
  {Efficient six-dimensional phase space reconstructions from experimental
  measurements using generative machine learning},\ }\href
  {https://doi.org/10.1103/PhysRevAccelBeams.27.094601} {\bibfield  {journal}
  {\bibinfo  {journal} {Phys. Rev. Accel. Beams}\ }\textbf {\bibinfo {volume}
  {27}},\ \bibinfo {pages} {094601} (\bibinfo {year} {2024})}\BibitemShut
  {NoStop}%
\bibitem [{\citenamefont {D'Arcy}\ \emph {et~al.}(2019)\citenamefont {D'Arcy}
  \emph {et~al.}}]{d2019flashforward}%
  \BibitemOpen
  \bibfield  {author} {\bibinfo {author} {\bibfnamefont {R.}~\bibnamefont
  {D'Arcy}} \emph {et~al.},\ }\bibfield  {title} {\bibinfo {title}
  {Flashforward: plasma wakefield accelerator science for high-average-power
  applications},\ }\href {https://doi.org/10.1098/rsta.2018.0392} {\bibfield
  {journal} {\bibinfo  {journal} {Phil. Trans. R. Soc. A}\ }\textbf {\bibinfo
  {volume} {377}},\ \bibinfo {pages} {20180392} (\bibinfo {year}
  {2019})}\BibitemShut {NoStop}%
\bibitem [{\citenamefont {Schreiber}\ \emph {et~al.}(2015)\citenamefont
  {Schreiber} \emph {et~al.}}]{schreiber_faatz_2015}%
  \BibitemOpen
  \bibfield  {author} {\bibinfo {author} {\bibfnamefont {S.}~\bibnamefont
  {Schreiber}} \emph {et~al.},\ }\bibfield  {title} {\bibinfo {title} {{The
  free-electron laser FLASH}},\ }\href {https://doi.org/10.1017/hpl.2015.16}
  {\bibfield  {journal} {\bibinfo  {journal} {High Power Laser Sci. Eng.}\
  }\textbf {\bibinfo {volume} {3}},\ \bibinfo {pages} {e20} (\bibinfo {year}
  {2015})}\BibitemShut {NoStop}%
\bibitem [{\citenamefont {Ackermann}\ \emph {et~al.}(2007)\citenamefont
  {Ackermann} \emph {et~al.}}]{Ackermann2007}%
  \BibitemOpen
  \bibfield  {author} {\bibinfo {author} {\bibfnamefont {W.}~\bibnamefont
  {Ackermann}} \emph {et~al.},\ }\bibfield  {title} {\bibinfo {title}
  {Operation of a free-electron laser from the extreme ultraviolet to the water
  window},\ }\href {https://doi.org/10.1038/nphoton.2007.76} {\bibfield
  {journal} {\bibinfo  {journal} {Nat. Photonics}\ }\textbf {\bibinfo {volume}
  {1}},\ \bibinfo {pages} {336} (\bibinfo {year} {2007})}\BibitemShut {NoStop}%
\bibitem [{\citenamefont {Wiebers}\ \emph {et~al.}(2013)\citenamefont {Wiebers}
  \emph {et~al.}}]{wiebers:ibic13-wepf03}%
  \BibitemOpen
  \bibfield  {author} {\bibinfo {author} {\bibfnamefont {C.}~\bibnamefont
  {Wiebers}} \emph {et~al.},\ }\bibfield  {title} {\bibinfo {title}
  {{Scintillating Screen Monitors for Transverse Electron Beam Profile
  Diagnostics at the European XFEL}},\ }in\ \href
  {https://jacow.org/IBIC2013/papers/WEPF03.pdf} {\emph {\bibinfo {booktitle}
  {Proc. IBIC'13}}}\ (\bibinfo  {publisher} {JACoW Publishing, Geneva,
  Switzerland},\ \bibinfo {year} {2013})\ pp.\ \bibinfo {pages}
  {807--810}\BibitemShut {NoStop}%
\bibitem [{\citenamefont {Courant}\ and\ \citenamefont
  {Snyder}(1958)}]{COURANT19581}%
  \BibitemOpen
  \bibfield  {author} {\bibinfo {author} {\bibfnamefont {E.}~\bibnamefont
  {Courant}}\ and\ \bibinfo {author} {\bibfnamefont {H.}~\bibnamefont
  {Snyder}},\ }\bibfield  {title} {\bibinfo {title} {Theory of the
  alternating-gradient synchrotron},\ }\href
  {https://doi.org/10.1016/0003-4916(58)90012-5} {\bibfield  {journal}
  {\bibinfo  {journal} {Annals of Physics}\ }\textbf {\bibinfo {volume} {3}},\
  \bibinfo {pages} {1} (\bibinfo {year} {1958})}\BibitemShut {NoStop}%
\bibitem [{\citenamefont {Andersen}\ and\ \citenamefont
  {Kak}(1984)}]{ANDERSEN198481}%
  \BibitemOpen
  \bibfield  {author} {\bibinfo {author} {\bibfnamefont {A.}~\bibnamefont
  {Andersen}}\ and\ \bibinfo {author} {\bibfnamefont {A.}~\bibnamefont {Kak}},\
  }\bibfield  {title} {\bibinfo {title} {{S}imultaneous {A}lgebraic
  {R}econstruction {T}echnique ({SART}): A superior implementation of the {ART}
  algorithm},\ }\href {https://doi.org/10.1016/0161-7346(84)90008-7} {\bibfield
   {journal} {\bibinfo  {journal} {Ultrason. Imaging}\ }\textbf {\bibinfo
  {volume} {6}},\ \bibinfo {pages} {81} (\bibinfo {year} {1984})}\BibitemShut
  {NoStop}%
\bibitem [{\citenamefont {Van~der Walt}\ \emph {et~al.}(2014)\citenamefont
  {Van~der Walt} \emph {et~al.}}]{van2014scikit}%
  \BibitemOpen
  \bibfield  {author} {\bibinfo {author} {\bibfnamefont {S.}~\bibnamefont
  {Van~der Walt}} \emph {et~al.},\ }\bibfield  {title} {\bibinfo {title}
  {scikit-image: image processing in {P}ython},\ }\href
  {https://doi.org/10.7717/peerj.453} {\bibfield  {journal} {\bibinfo
  {journal} {PeerJ}\ }\textbf {\bibinfo {volume} {2}},\ \bibinfo {pages} {e453}
  (\bibinfo {year} {2014})}\BibitemShut {NoStop}%
\bibitem [{\citenamefont {Derbenev}\ \emph {et~al.}(1995)\citenamefont
  {Derbenev}, \citenamefont {Rossbach}, \citenamefont {Saldin},\ and\
  \citenamefont {Shiltsev}}]{Derbenev:414678}%
  \BibitemOpen
  \bibfield  {author} {\bibinfo {author} {\bibfnamefont {Y.~S.}\ \bibnamefont
  {Derbenev}}, \bibinfo {author} {\bibfnamefont {J.}~\bibnamefont {Rossbach}},
  \bibinfo {author} {\bibfnamefont {E.~L.}\ \bibnamefont {Saldin}},\ and\
  \bibinfo {author} {\bibfnamefont {V.~D.}\ \bibnamefont {Shiltsev}},\ }\href
  {https://doi.org/10.3204/PUBDB-2018-04128} {\emph {\bibinfo {title}
  {{M}icrobunch radiative tail - head interaction}}},\ \bibinfo {type} {Tech.
  Rep.}\ \bibinfo {number} {TESLA-FEL 1995-05}\ (\bibinfo  {institution}
  {DESY},\ \bibinfo {year} {1995})\BibitemShut {NoStop}%
\bibitem [{\citenamefont {Dohlus}\ \emph {et~al.}(2006)\citenamefont {Dohlus}
  \emph {et~al.}}]{dohlus2006bunch}%
  \BibitemOpen
  \bibfield  {author} {\bibinfo {author} {\bibfnamefont {M.}~\bibnamefont
  {Dohlus}} \emph {et~al.},\ }\bibfield  {title} {\bibinfo {title} {{Bunch
  Compression for Linac-Based {FEL}'s. Electron Bunch Length Compression}},\
  }\href {https://www.osti.gov/biblio/877227} {\bibfield  {journal} {\bibinfo
  {journal} {ICFA Beam Dynamics Newsletter}\ }\textbf {\bibinfo {volume} {38}}
  (\bibinfo {year} {2006})}\BibitemShut {NoStop}%
\bibitem [{\citenamefont {Rivkin}(1986)}]{rivkin1986damping}%
  \BibitemOpen
  \bibfield  {author} {\bibinfo {author} {\bibfnamefont {L.~Z.}\ \bibnamefont
  {Rivkin}},\ }\emph {\bibinfo {title} {Damping ring for the SLAC linear
  collider}},\ \href
  {https://lib-extopc.kek.jp/preprints/PDF/1986/8603/8603010.pdf} {Ph.D.
  thesis},\ \bibinfo  {school} {California Institute of Technology} (\bibinfo
  {year} {1986})\BibitemShut {NoStop}%
\bibitem [{\citenamefont {Prat}\ and\ \citenamefont
  {Aiba}(2014)}]{PhysRevSTAB.17.052801}%
  \BibitemOpen
  \bibfield  {author} {\bibinfo {author} {\bibfnamefont {E.}~\bibnamefont
  {Prat}}\ and\ \bibinfo {author} {\bibfnamefont {M.}~\bibnamefont {Aiba}},\
  }\bibfield  {title} {\bibinfo {title} {Four-dimensional transverse beam
  matrix measurement using the multiple-quadrupole scan technique},\ }\href
  {https://doi.org/10.1103/PhysRevSTAB.17.052801} {\bibfield  {journal}
  {\bibinfo  {journal} {Phys. Rev. ST Accel. Beams}\ }\textbf {\bibinfo
  {volume} {17}},\ \bibinfo {pages} {052801} (\bibinfo {year}
  {2014})}\BibitemShut {NoStop}%
\bibitem [{\citenamefont {Aicheler}\ \emph {et~al.}(2012)\citenamefont
  {Aicheler}, \citenamefont {Burrows}, \citenamefont {Draper}, \citenamefont
  {Garvey}, \citenamefont {Lebrun}, \citenamefont {Peach}, \citenamefont
  {Phinney}, \citenamefont {Schmickler}, \citenamefont {Schulte},\ and\
  \citenamefont {Toge}}]{Aicheler:1500095}%
  \BibitemOpen
  \bibfield  {author} {\bibinfo {author} {\bibfnamefont {M.}~\bibnamefont
  {Aicheler}}, \bibinfo {author} {\bibfnamefont {P.}~\bibnamefont {Burrows}},
  \bibinfo {author} {\bibfnamefont {M.}~\bibnamefont {Draper}}, \bibinfo
  {author} {\bibfnamefont {T.}~\bibnamefont {Garvey}}, \bibinfo {author}
  {\bibfnamefont {P.}~\bibnamefont {Lebrun}}, \bibinfo {author} {\bibfnamefont
  {K.}~\bibnamefont {Peach}}, \bibinfo {author} {\bibfnamefont
  {N.}~\bibnamefont {Phinney}}, \bibinfo {author} {\bibfnamefont
  {H.}~\bibnamefont {Schmickler}}, \bibinfo {author} {\bibfnamefont
  {D.}~\bibnamefont {Schulte}},\ and\ \bibinfo {author} {\bibfnamefont
  {N.}~\bibnamefont {Toge}},\ }\href {https://doi.org/10.5170/CERN-2012-007}
  {\emph {\bibinfo {title} {{A Multi-TeV Linear Collider Based on CLIC
  Technology: CLIC Conceptual Design Report}}}},\ CERN Yellow Reports:
  Monographs\ (\bibinfo  {publisher} {CERN},\ \bibinfo {address} {Geneva},\
  \bibinfo {year} {2012})\BibitemShut {NoStop}%
\bibitem [{\citenamefont {Ferran~Pousa}\ \emph {et~al.}(2023)\citenamefont
  {Ferran~Pousa}, \citenamefont {Jalas}, \citenamefont {Kirchen}, \citenamefont
  {Martinez de~la Ossa}, \citenamefont {Th\'evenet}, \citenamefont {Hudson},
  \citenamefont {Larson}, \citenamefont {Huebl}, \citenamefont {Vay},\ and\
  \citenamefont {Lehe}}]{PhysRevAccelBeams.26.084601}%
  \BibitemOpen
  \bibfield  {author} {\bibinfo {author} {\bibfnamefont {A.}~\bibnamefont
  {Ferran~Pousa}}, \bibinfo {author} {\bibfnamefont {S.}~\bibnamefont {Jalas}},
  \bibinfo {author} {\bibfnamefont {M.}~\bibnamefont {Kirchen}}, \bibinfo
  {author} {\bibfnamefont {A.}~\bibnamefont {Martinez de~la Ossa}}, \bibinfo
  {author} {\bibfnamefont {M.}~\bibnamefont {Th\'evenet}}, \bibinfo {author}
  {\bibfnamefont {S.}~\bibnamefont {Hudson}}, \bibinfo {author} {\bibfnamefont
  {J.}~\bibnamefont {Larson}}, \bibinfo {author} {\bibfnamefont
  {A.}~\bibnamefont {Huebl}}, \bibinfo {author} {\bibfnamefont {J.-L.}\
  \bibnamefont {Vay}},\ and\ \bibinfo {author} {\bibfnamefont {R.}~\bibnamefont
  {Lehe}},\ }\bibfield  {title} {\bibinfo {title} {Bayesian optimization of
  laser-plasma accelerators assisted by reduced physical models},\ }\href
  {https://doi.org/10.1103/PhysRevAccelBeams.26.084601} {\bibfield  {journal}
  {\bibinfo  {journal} {Phys. Rev. Accel. Beams}\ }\textbf {\bibinfo {volume}
  {26}},\ \bibinfo {pages} {084601} (\bibinfo {year} {2023})}\BibitemShut
  {NoStop}%
\bibitem [{\citenamefont {Emma}\ \emph {et~al.}(2002)\citenamefont {Emma} \emph
  {et~al.}}]{emma:epac02-thpri097}%
  \BibitemOpen
  \bibfield  {author} {\bibinfo {author} {\bibfnamefont {P.}~\bibnamefont
  {Emma}} \emph {et~al.},\ }\bibfield  {title} {\bibinfo {title} {{Bunch Length
  Measurements Using a Transverse RF Deflecting Structure in the SLAC Linac}},\
  }in\ \href {https://jacow.org/e02/papers/THPRI097.pdf} {\emph {\bibinfo
  {booktitle} {Proc. EPAC'02}}}\ (\bibinfo  {publisher} {JACoW Publishing,
  Geneva, Switzerland},\ \bibinfo {year} {2002})\ pp.\ \bibinfo {pages}
  {1882--1884}\BibitemShut {NoStop}%
\bibitem [{\citenamefont {Behrens}\ \emph {et~al.}(2014)\citenamefont {Behrens}
  \emph {et~al.}}]{Behrens2014}%
  \BibitemOpen
  \bibfield  {author} {\bibinfo {author} {\bibfnamefont {C.}~\bibnamefont
  {Behrens}} \emph {et~al.},\ }\bibfield  {title} {\bibinfo {title}
  {Few-femtosecond time-resolved measurements of x-ray free-electron lasers},\
  }\href {https://doi.org/10.1038/ncomms4762} {\bibfield  {journal} {\bibinfo
  {journal} {Nature Communications}\ }\textbf {\bibinfo {volume} {5}},\
  \bibinfo {pages} {3762} (\bibinfo {year} {2014})}\BibitemShut {NoStop}%
\bibitem [{\citenamefont {Huang}\ \emph {et~al.}(2007)\citenamefont {Huang},
  \citenamefont {Lu}, \citenamefont {Zhou}, \citenamefont {Clayton},
  \citenamefont {Joshi}, \citenamefont {Mori}, \citenamefont {Muggli},
  \citenamefont {Deng}, \citenamefont {Oz}, \citenamefont {Katsouleas},
  \citenamefont {Hogan}, \citenamefont {Blumenfeld}, \citenamefont {Decker},
  \citenamefont {Ischebeck}, \citenamefont {Iverson}, \citenamefont {Kirby},\
  and\ \citenamefont {Walz}}]{PhysRevLett.99.255001}%
  \BibitemOpen
  \bibfield  {author} {\bibinfo {author} {\bibfnamefont {C.}~\bibnamefont
  {Huang}}, \bibinfo {author} {\bibfnamefont {W.}~\bibnamefont {Lu}}, \bibinfo
  {author} {\bibfnamefont {M.}~\bibnamefont {Zhou}}, \bibinfo {author}
  {\bibfnamefont {C.~E.}\ \bibnamefont {Clayton}}, \bibinfo {author}
  {\bibfnamefont {C.}~\bibnamefont {Joshi}}, \bibinfo {author} {\bibfnamefont
  {W.~B.}\ \bibnamefont {Mori}}, \bibinfo {author} {\bibfnamefont
  {P.}~\bibnamefont {Muggli}}, \bibinfo {author} {\bibfnamefont
  {S.}~\bibnamefont {Deng}}, \bibinfo {author} {\bibfnamefont {E.}~\bibnamefont
  {Oz}}, \bibinfo {author} {\bibfnamefont {T.}~\bibnamefont {Katsouleas}},
  \bibinfo {author} {\bibfnamefont {M.~J.}\ \bibnamefont {Hogan}}, \bibinfo
  {author} {\bibfnamefont {I.}~\bibnamefont {Blumenfeld}}, \bibinfo {author}
  {\bibfnamefont {F.~J.}\ \bibnamefont {Decker}}, \bibinfo {author}
  {\bibfnamefont {R.}~\bibnamefont {Ischebeck}}, \bibinfo {author}
  {\bibfnamefont {R.~H.}\ \bibnamefont {Iverson}}, \bibinfo {author}
  {\bibfnamefont {N.~A.}\ \bibnamefont {Kirby}},\ and\ \bibinfo {author}
  {\bibfnamefont {D.}~\bibnamefont {Walz}},\ }\bibfield  {title} {\bibinfo
  {title} {Hosing instability in the blow-out regime for plasma-wakefield
  acceleration},\ }\href {https://doi.org/10.1103/PhysRevLett.99.255001}
  {\bibfield  {journal} {\bibinfo  {journal} {Phys. Rev. Lett.}\ }\textbf
  {\bibinfo {volume} {99}},\ \bibinfo {pages} {255001} (\bibinfo {year}
  {2007})}\BibitemShut {NoStop}%
\bibitem [{\citenamefont {Mehrling}\ \emph {et~al.}(2017)\citenamefont
  {Mehrling}, \citenamefont {Fonseca}, \citenamefont {Martinez de~la Ossa},\
  and\ \citenamefont {Vieira}}]{PhysRevLett.118.174801}%
  \BibitemOpen
  \bibfield  {author} {\bibinfo {author} {\bibfnamefont {T.~J.}\ \bibnamefont
  {Mehrling}}, \bibinfo {author} {\bibfnamefont {R.~A.}\ \bibnamefont
  {Fonseca}}, \bibinfo {author} {\bibfnamefont {A.}~\bibnamefont {Martinez
  de~la Ossa}},\ and\ \bibinfo {author} {\bibfnamefont {J.}~\bibnamefont
  {Vieira}},\ }\bibfield  {title} {\bibinfo {title} {Mitigation of the hose
  instability in plasma-wakefield accelerators},\ }\href
  {https://doi.org/10.1103/PhysRevLett.118.174801} {\bibfield  {journal}
  {\bibinfo  {journal} {Phys. Rev. Lett.}\ }\textbf {\bibinfo {volume} {118}},\
  \bibinfo {pages} {174801} (\bibinfo {year} {2017})}\BibitemShut {NoStop}%
\bibitem [{\citenamefont {Burkart}\ \emph {et~al.}(2022)\citenamefont {Burkart}
  \emph {et~al.}}]{burkart:linac22-thpojo01}%
  \BibitemOpen
  \bibfield  {author} {\bibinfo {author} {\bibfnamefont {F.}~\bibnamefont
  {Burkart}} \emph {et~al.},\ }\bibfield  {title} {\bibinfo {title} {{The ARES
  Linac at DESY}},\ }in\ \href
  {https://doi.org/10.18429/JACoW-LINAC2022-THPOJO01} {\emph {\bibinfo
  {booktitle} {Proc. LINAC'22}}},\ \bibinfo {series and number} {\bibinfo
  {series} {International Linear Accelerator Conference}\ No.~\bibinfo {number}
  {31}}\ (\bibinfo  {publisher} {JACoW Publishing, Geneva, Switzerland},\
  \bibinfo {year} {2022})\ pp.\ \bibinfo {pages} {691--694}\BibitemShut
  {NoStop}%
\bibitem [{\citenamefont {Hoover}\ and\ \citenamefont
  {Wong}(2024)}]{PhysRevResearch.6.033163}%
  \BibitemOpen
  \bibfield  {author} {\bibinfo {author} {\bibfnamefont {A.}~\bibnamefont
  {Hoover}}\ and\ \bibinfo {author} {\bibfnamefont {J.~C.}\ \bibnamefont
  {Wong}},\ }\bibfield  {title} {\bibinfo {title} {High-dimensional
  maximum-entropy phase space tomography using normalizing flows},\ }\href
  {https://doi.org/10.1103/PhysRevResearch.6.033163} {\bibfield  {journal}
  {\bibinfo  {journal} {Phys. Rev. Res.}\ }\textbf {\bibinfo {volume} {6}},\
  \bibinfo {pages} {033163} (\bibinfo {year} {2024})}\BibitemShut {NoStop}%
\bibitem [{\citenamefont
  {Hoover}(2025)}]{hoover2025ndimensionalmaximumentropytomographyparticle}%
  \BibitemOpen
  \bibfield  {author} {\bibinfo {author} {\bibfnamefont {A.}~\bibnamefont
  {Hoover}},\ }\href {https://arxiv.org/abs/2409.17915} {\bibinfo {title}
  {N-dimensional maximum-entropy tomography via particle sampling}} (\bibinfo
  {year} {2025}),\ \Eprint {https://arxiv.org/abs/2409.17915} {arXiv:2409.17915
  [physics.acc-ph]} \BibitemShut {NoStop}%
\end{thebibliography}%

\end{document}